\documentclass[12pt]{iopart}
\pdfoutput=1
\usepackage{graphicx}
\begin{document}

\title[Detailed Transport Investigation of the Magnetic Anisotropy of (Ga,Mn)As]{Detailed Transport Investigation of the Magnetic Anisotropy of (Ga,Mn)As}

\author{K Pappert, C Gould,}

\address{Physikalisches Institut (EP3), Universit\"{a}t W\"{u}rzburg, Am
Hubland, D-97074 W\"{u}rzburg, Germany}

\author{M Sawicki,}
\address{Institute of Physics, Polish Academy of Sciences, al. Lotnik\'{o}w 32/46, PL-02668, Warszawa, Poland}

\author{J Wenisch, K Brunner, G Schmidt, L W Molenkamp}
\address{Physikalisches Institut (EP3), Universit\"{a}t W\"{u}rzburg, Am
Hubland, D-97074 W\"{u}rzburg, Germany}

\ead{gould@physik.uni-wuerzburg.de}

\begin{abstract}

This paper discusses transport methods for the investigation of the
(Ga,Mn)As magnetic anisotropy. Typical magnetoresistance behaviour
for different anisotropy types is discussed, focusing on an in depth
discussion of the anisotropy fingerprint technique and extending it
to layers with primarily uniaxial magnetic anisotropy.

We find that in all (Ga,Mn)As films studied, three anisotropy
components are always present; The primary biaxial along ([100] and
[010]) along with both uniaxial components along the
[$\overline{1}$10] and [010] crystal direction which are often
reported separately. Various fingerprints of typical (Ga,Mn)As
transport samples at 4~K are included to illustrate the variation of
the relative strength of these anisotropy terms. We further
investigate the temperature dependence of the magnetic anisotropy
and the domain wall nucleation energy with the help of the
fingerprint method.

\end{abstract}

\maketitle

As the sophistication of spintronic device investigations
continues to rapidly grow, a deeper and more detailed
characterization of the ferromagnetic semiconductor material used
in the elaboration of many of these structures is becoming ever
more essential to properly understanding the operation and design
of device elements. The spin-orbit mediated coupling of magnetic
and semiconductor properties in these materials gives rise to many
novel transport-related phenomena which can be harnessed for
device applications. For the understanding and reliable
functioning of such devices it is important to understand and be
able to determine the magnetic anisotropy of the parent layer and
of readily structured samples. While FMR \cite{Furdyna} and SQUID
\cite{Mike05PRL} can effectively measure the main magnetic
anisotropy of the parent layer, they are not practical for
anisotropy studies on individual small structures which have too
little magnetic moment to be detected. Transport measurements, on
the other hand provide a very effective means of extensively
studying the anisotropy at a fixed temperature. Using a vector
field magnet, many magnetic field scans in different in-plane (or
even space-) directions can be recorded within a short time frame
without remounting the sample. Anisotropic transport properties
allow for electrical monitoring of the magnetization. This
provides detailed information on the angular dependence of the
magnetic behaviour.

A technique for extracting the magnetic anisotropy by transport
means was introduced in \cite{fingerprint}. In this treatise we
discuss investigations of the magnetization behaviour by tranport
means in general and in particular the anisotropy fingerprint
technique in much greater detail. We present a variety of
fingerprints of different (Ga,Mn)As layers at 4~K and discuss the
always present three symmetry components of the magnetic
anisotropy at 4~K. We then extend the method to the case of a
uniaxial material, which is necessary to describe (Ga,Mn)As layers
at higher temperatures or structured submicron devices. We
investigate the temperature behaviour of the (Ga,Mn)As anisotropy
using the fingerprint method. It shows the typical transition from
a mainly biaxial system at low temperature to a uniaxial system
close to T$_{C}$. From these fingerprints we can also extract the
temperature dependence of the domain wall nucleation and
propagation energy.

\section{Anisotropic Transport and Magnetic Anisotropy in (Ga,Mn)As}\label{sec:aniso}

The ferromagnetic semiconductor (Ga,Mn)As is strongly anisotropic
both in transport and in its magnetic properties. It shows a
strong anisotropic magnetoresistance effect (AMR): The resistivity
for a current flow perpendicular to the magnetization
$\rho_{\bot}$ is larger than $\rho_{||}$ parallel to the
magnetization\cite{Baxter}. Ohm's law is best expressed with the
electric field \textbf{E} broken up in components parallel and
perpendicular to the magnetization \textbf{M}
\cite{Jan1957,McGuire}

\begin{center}
{\parbox{15cm}{
\begin{equation}
\label{eqn:AMR}
\mathbf{E}=\rho_{||}\mathbf{J_{||}}+\rho_{\bot}\mathbf{J_{\bot}}
\end{equation}
}}
\end{center}

with \textbf{J} the current density. The projection onto the
current path gives the longitudinal resistivity $\rho_{xx}$
(longitudinal AMR effect):

\begin{center}
{\parbox{15cm}{
\begin{equation}
\label{eqn:Rxx}
\rho_{xx}=\rho_{\bot}-(\rho_{\bot}-\rho_{||})\cos^2(\vartheta),
\end{equation}
}}
\end{center}

where $\vartheta$ is the angle between \textbf{M} and \textbf{J}.
The dependence of the Hall resistivity $\rho_{xy}$ (transverse AMR
or Planar Hall effect PHE) on the magnetization direction follows
directly from the electric field component perpendicular to the
current path:

\begin{center}
{\parbox{15cm}{
\begin{equation}
\label{eqn:Rxy}
\rho_{xy}=-\frac{\rho_{\bot}-\rho_{||}}{2}\sin(2\vartheta),
\end{equation}
}}
\end{center}

With the help of longitudinal AMR and PHE measurements it is thus
possible to monitor the magnetization direction $\vartheta$ and
conclude on the magnetic anisotropy of the material.

The cubic anisotropy of the crystal structure is reduced by growth
strain. Here we discuss highly doped not annealed (Ga,Mn)As layers
grown under compressive strain on GaAs (001) substrates. For
standard thickness and experimentally relevant hole densities, the
growth strain results in
an additional strong hard axis in growth direction that confines
the easy axes to the layer plane. This in-plane anisotropy is
strongly temperature dependent as will be discussed in section
\ref{ch:T-dep}. At 4~K the material shows a main biaxial magnetic
anisotropy with easy axes along the [100] and [010] crystal
direction. The above is well understood, however, in addition to
this two uniaxial anisotropy terms have been observed the origin
of which is not clear. One additional uniaxial anisotropy term
with easy axis along [110] or $[\overline{1}10]$ is typically
present and has been seen in many laboratories. A much smaller
additional uniaxial anisotropy component with easy axis along
[010] or [100] \cite{Chris1} has often been overlooked, because it
is typically too small to be visible in standard SQUID
measurements. Recently, the anisotropy fingerprint technique
\cite{fingerprint} allowed us to show that all three, the main
biaxial and the two uniaxial, anisotropy components are
simultaneously present in typical (Ga,Mn)As layers at 4~K. Section
\ref{sec:fingerprints} explains the details of the method.
Fingerprints of typical (Ga,Mn)As layers are shown in section
\ref{ssec:FPGallery} to discuss the typical relative strength of
the anisotropy components and their variation from layer to layer.

In this context it is helpful to note that for the purpose of
calculating the magnetostatic energy in the single domain model,
any linear combination of uniaxial anisotropy components with
different easy axes can be expressed as a linear combination of a
[$\overline{1}$10] and a [010] uniaxial anisotropy term. It is
known that:

\begin{center}
{\parbox{15cm}{
\begin{equation}
\label{eqn:sines} a\sin\alpha+b\cos\alpha=\sqrt{a^2+b^2} \cdot
\sin (\alpha+\beta),
\end{equation}
}}
\end{center}

where $\beta$ is given by $\arctan(b/a)$ and $\arctan(b/a)\pm\pi$
if $a\geq0$ and $a<0$ respectively. This relates two sine
functions of the same period but with different phase to a third
sine function with the same period and a new phase. Consequently,
we can express any combination of two uniaxial anisotropy
components in a (Ga,Mn)As layer by an equivalent linear
combination of the [$\overline{1}$10] and the [010] uniaxial
anisotropy term. The choice of only these two directions is thus
fully general and does not exclude other uniaxial anisotropy
components, e.g. due to specific strain conditions in a specific
sample.

Summing up the three anisotropy terms of different symmetry, we
can express the magnetostatic energy E of a magnetic domain with
magnetization orientation $\vartheta$ with respect to the
[100]-crystal direction:

\begin{center}
{\parbox{15cm}{
\begin{equation}
\label{eqn:E} E = \frac{K_{cryst}}{4}\sin^{2}(2 \vartheta)
+K_{uni[\overline{1}10]}\sin^{2}(\vartheta-135^\circ)+K_{uni[010]}\sin^{2}(\vartheta-90^\circ)
-MH\cos(\vartheta-\varphi),
\end{equation}
}}
\end{center}

where the last term is the Zeeman energy. The anisotropy constants
$K_{cryst}$ in the biaxial anisotropy term and
$K_{uni[\overline{1}10]}$ and $K_{uni[010]}$ in the two uniaxial
terms depend differently on the magnetization \textbf{M} and thus
on temperature\cite{Mike05PRL}. This results in a characteristic
temperature dependence of the overall magnetic anisotropy of the
layer. This typical transition from mainly biaxial behaviour at
4~K to uniaxial behaviour close to the Curie temperature is
investigated with the anisotropy fingerprint technique in section
\ref{sec:fingerprints}.

\begin{figure}[t]
\centering
\includegraphics[width=10cm]{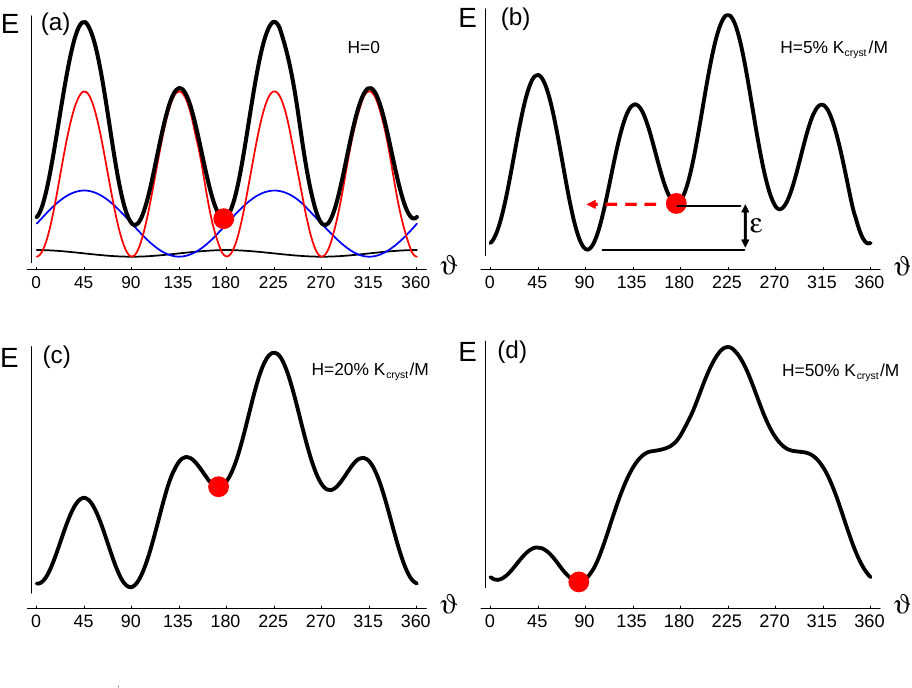}
\caption{\label{fig:Elandscape} \textit{Energy landscape at zero
field (a). The symmetry components of the anisotropy are shown
with thin lines (biaxial red; uniaxial along $[\overline{1}10]$
blue; uniaxial along [010] black). The energy surface evolves with
increasing field along $45^\circ$ (b-d) causing magnetization
reversal through domain wall nucleation and propagation (b) or
through Stoner-Wohlfarth rotation (c and d).} }
\end{figure}

Fig.~\ref{fig:Elandscape} shows the energy landscape, a plot of
the energy of a magnetic domain as a function of the magnetization
angle, and how it evolves with magnetic field. Under an applied
magnetic field, the magnetization can reverse through two
mechanisms. One mechanism is called coherent (Stoner-Wohlfarth
\cite{StonerWohlf}) rotation: With increasing magnetic field the
magnetization (marked with a red dot in Fig.~\ref{fig:Elandscape})
follows the local minimum of the energy surface until the minimum
disappears as illustrated in panel b to d. The dashed arrow in
panel b, on the other hand, illustrates the reversal by DW
nucleation and propagation. It appears if the energy gained by
reorienting the magnetization direction to another local minimum
of the energy surface is larger than the DW nucleation/propagation
energy $\varepsilon$. A DW is nucleated and a new domain with the
new magnetization orientation grows until it extends over the
whole structure. As will become evident, the inherent behaviour of
(Ga,Mn)As is generally dominated by Stoner-Wohlfarth rotation at
high magnetic fields and by DW-nucleation/propagation related
events at low fields.

\section{Monitoring the Magnetization Behaviour in Transport}\label{sec:el-biaxial}

The described magnetization behaviour can be observed in direct or
indirect magnetization measurements, and leads to a very
characteristic two-step reversal process in SQUID and
magnetoresistance measurements. Three-jump magnetic switching is
also possible in very specific situations\cite{Cowburn3jump}.

\begin{figure}[t]
\centering
\includegraphics[width=6cm]{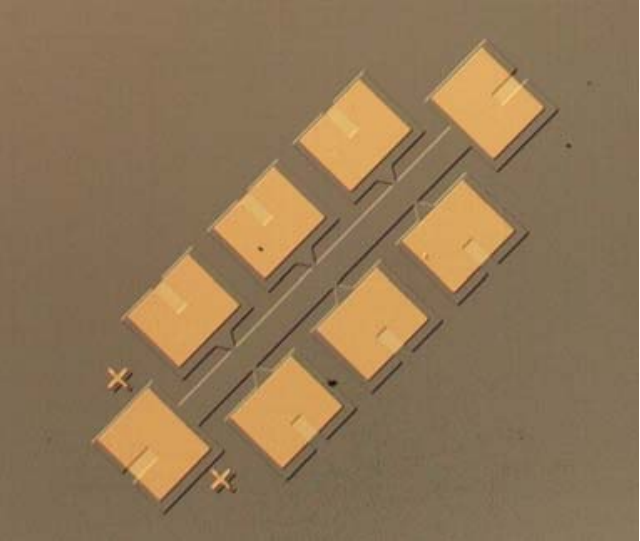}
\caption{\label{fig:Hallbar} \textit{Hall bar structure typical of
those used in this study, processed by optical lithography and dry
etching. The contacts are established by gold deposition. This bar
is 40 $\mu$m wide and 540 $\mu$m long.} }
\end{figure}

\begin{figure}[t] \centering
\includegraphics{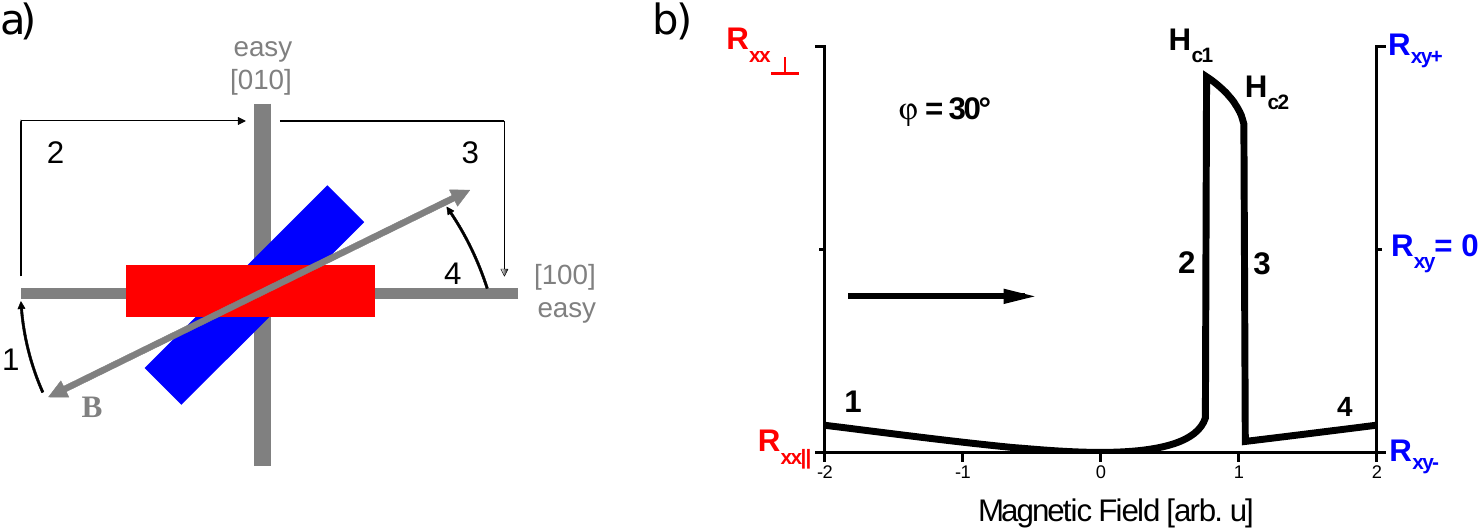}
\caption{\label{fig:AMRsim} \textit{ Two-step magnetization
reversal. a) Sketch showing the magnetization behaviour in hard
(blue) and easy (red) axis Hall bars. b) The corresponding
calculated AMR scan for the easy axis Hall bar (left scale), which
is equivalent to a Planar Hall scan on the hard axis bar (right
scale).} }
\end{figure}

Here we will discuss the characterization of the magnetic
anisotropy of typical Ga$_{1-x}$Mn$_{x}$As transport layers. The
layers were grown by low-temperature molecular beam epitaxy
(LT-MBE, 270$^\circ$C) on a high-quality GaAs buffer on an
epiready semiinsulating GaAs (001) substrate. They contain between
x=2\% and 5\% Mn and show an as-grown T$_C$ around 50~K or above.
All layers were patterned into 40 to 60 $\mu$m wide Hall bar
structures as shown in Fig.~\ref{fig:Hallbar} by optical
lithography and chlorine assisted dry etching. Contacts are
established through metal evaporation and lift off. During the
processing care is taken to not expose the samples to any
annealing treatment.

Assume a biaxial magnetic anisotropy with easy axes along the
in-plane $\langle100\rangle$ crystal directions (coordinate axes
in Fig.~\ref{fig:AMRsim}a) as a first approximation of the 4~K
anisotropy of (Ga,Mn)As. Assume further that the longitudinal
resistance of a Hall bar with its current along the [100] axis is
measured while the external magnetic field is swept from a high
negative to a high positive value along a direction 30$^\circ$
away from the [100] axis. Using eq.~\ref{eqn:E} and \ref{eqn:Rxx}
we can now calculate the corresponding AMR signal shown in
Fig.~\ref{fig:AMRsim}b(left y-axis scale). At high negative
fields, the magnetization is forced along the field direction (not
shown). (1) As the field is decreased \textbf{M} gradually relaxes
through Stoner-Wohlfarth rotation until it is aligned with its
closest easy axis. At zero field \textbf{M} is thus parallel to
$[\overline{1}00]$ and to the current, yielding the smallest
resistance value $R_{||}$. (2) At a small positive field H$_{c1}$
a 90$^\circ$-DW is nucleated and propagates through the structure
resulting in an abrupt change of the magnetization direction to
the [010] direction. \textbf{M} is now perpendicular to the
current, yielding the maximum resistance value R$_{\bot}$. (3) At
the second switching field H$_{c2}$, another 90$^\circ$-DW is
nucleated and the magnetization jumps close to the [100] easy
axis. (4) With increasing magnetic fields \textbf{M} rotates again
towards the magnetic field direction. The entire process is of
course hysteretically symmetric (not shown).

If another Hall bar is oriented along the [110] crystal direction
(blue in Fig.~\ref{fig:AMRsim}a) the easy axes [100] and [010]
have an angle of $\pm45^\circ$ with the current path. An abrupt
switch of magnetization from one easy axis to the other
corresponds according to eq.~\ref{eqn:Rxy} to a sharp switching
event between two extrema of the transverse resistance. The
calculated Planar Hall signal is thus up to a constant identical
with the previously discussed curve in Fig.~\ref{fig:AMRsim}b, in
this case centered around zero transverse resistance (blue/right
y-axis). Because of this, transverse resistance measurements are
the method of choice for Hall bars oriented along a crystalline
hard axis. For Hall bars along an easy axis, longitudinal
resistance measurements are the only useful technique. Indeed, if
the current direction is rotated by 45$^\circ$, Eq.~\ref{eqn:Rxy}
transforms into Eq.~\ref{eqn:Rxx} (plus an uninteresting offset).

\begin{figure}[t]
\includegraphics[width=16cm]{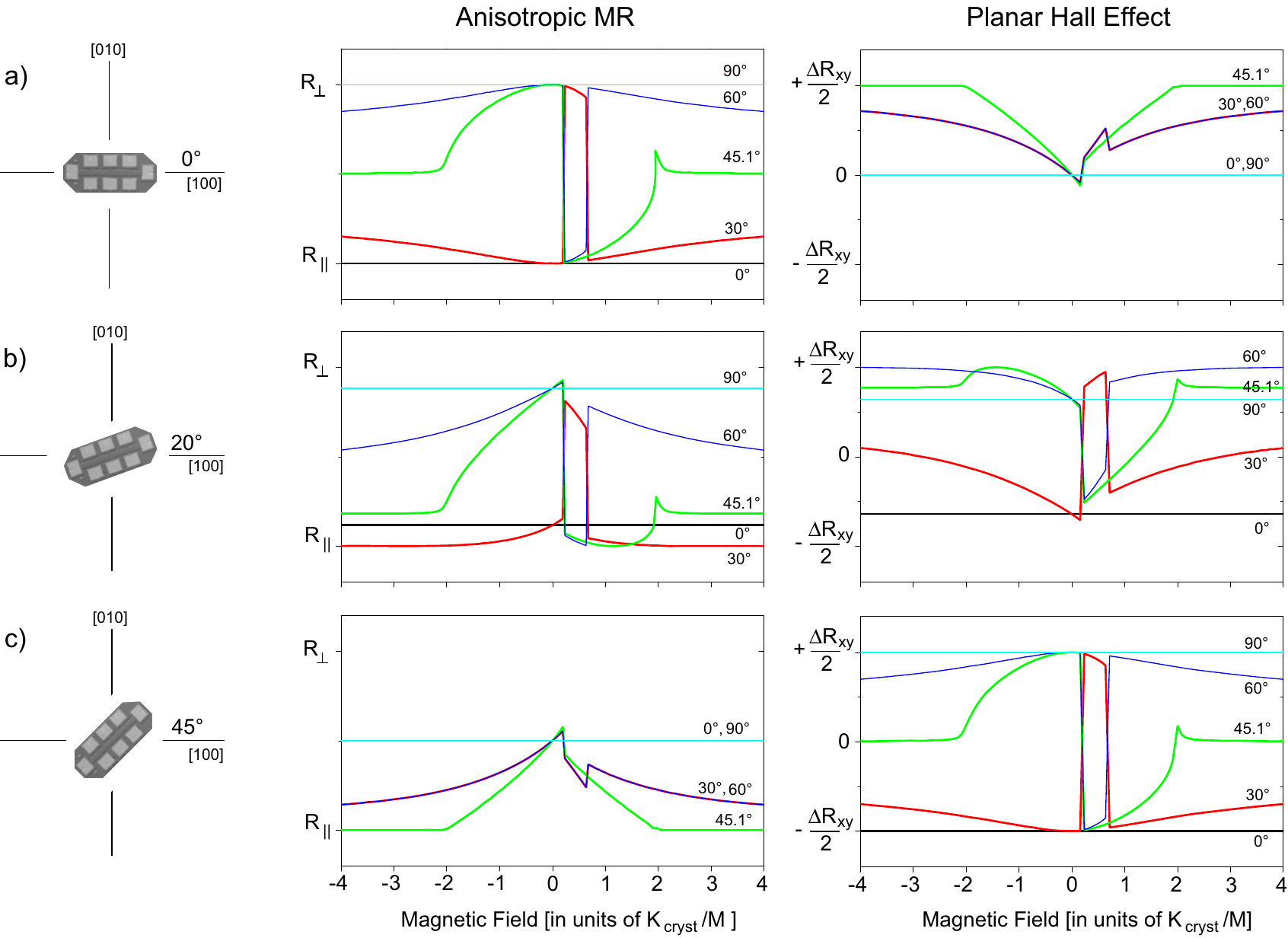}
\caption{\label{fig:AMR-IPH-panels} \textit{Calculated Anisotropic
Magnetoresistance (middle) and Planar Hall effect (right) curves
for magnetic field sweeps along several in-plane angles
($\varphi$= 0$^\circ$, 30$^\circ$, 45.1$^\circ$, 60$^\circ$ and
90$^\circ$) for Hall bar orientations as indicated in the sketches
on the left, with current along a) 0$^\circ$ b) 20$^\circ$ c)
45$^\circ$. The underlying magnetic anisotropy is biaxial with
easy axes along [100] and [010]. All angles with respect to the
[100] crystal direction. The domain wall nucleation/propagation
energy $\varepsilon$ is exaggerated with 30\% of $K_{cryst}$. } }
\end{figure}

Fig.~\ref{fig:AMR-IPH-panels} shows AMR (middle) and Planar Hall
effect (right) curves for field sweeps along different angles
$\varphi$ in the plane calculated using Eq.~\ref{eqn:E} in
combination with Eq.~\ref{eqn:Rxx} and~\ref{eqn:Rxy} respectively.
The domain wall nucleation energy $\varepsilon$ was exaggerated in
these calculations (30\% of $K_{cryst}$ instead of 5-10\% as would
be typical for (Ga,Mn)As)) to illustrate both Stoner-Wohlfarth
rotation and DW-related magnetization switching in the same
figure. The middle panel of Fig.~\ref{fig:AMR-IPH-panels}a, shows
MR curves for a Hall bar along a biaxial easy axis. If the
external magnetic field is swept along the [100] easy axis
(0$^\circ$), the magnetization is always parallel to the current
direction. The resistance (black line) thus takes its lowest value
R$_{||}$ throughout the whole field range. If the field is swept
along the [010] easy axis (90$^\circ$), the magnetization is
always perpendicular to the current resulting in a high resistance
value R$_{\bot}$ throughout the whole curve (thin cyan). For
intermediate magnetic field angles, the magnetization is parallel
to the field at high positive and negative fields, yielding
intermediate resistance values. At zero field the magnetization
relaxes to the closest easy axis, which is [100] for the
30$^\circ$ scan and [010] for the 60$^\circ$ and 45.1$^\circ$
scans, corresponding to the lowest and highest resistance value
respectively. The 45.1$^\circ$-scan (green line) can be used to
measure the strength of the magnetic anisotropy. We can read out
the anisotropy field (-2$K_{cryst}/M$), at the point where the
magnetization starts to turn away from the magnetic field
direction. A measurement with two possible resistance states at
zero field always suggests a biaxial magnetic anisotropy. However,
note that these two states can correspond to the same resistance
value as, e.g., if the easy axis and the current include an angle
of 45$^\circ$ (left panel of Fig.~\ref{fig:AMR-IPH-panels}c),
where the 0$^\circ$(black) and $90^\circ$(cyan) curve fall on top
of each other. The panels on the right show the calculated Planar
Hall resistance curves in the respective configurations. Note,
that, the AMR signal in Fig.~\ref{fig:AMR-IPH-panels}a is
identical to the PHE signal in Fig.~\ref{fig:AMR-IPH-panels}c, as
discussed above. The easy axis showing constant resistance
throughout the whole scan can easily be identified in any of the
configurations, even if current and easy axis include an oblique
angle as in Fig.~\ref{fig:AMR-IPH-panels}b.

\begin{figure}[t]
\centering
\includegraphics[width=8cm]{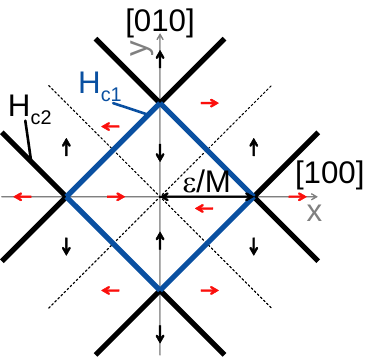}
\caption{\label{fig:polarLinesBiaxial} \textit{Switching field
positions (thick solid lines) in a polar plot for a biaxial
material with easy axes along [100] and [010](coordinate axes).
The magnetization direction in each region of the plot is
indicated by arrows (red/black: high/low resistance) and the hard
axes by dashed lines. } }
\end{figure}

 The switching fields (H$_{c1}$ and H$_{c2}$ in
Fig.~\ref{fig:AMRsim}b) can be derived analytically from
eq.~\ref{eqn:E} \cite{Cowb95} (here for a pure biaxial anisotropy;
$K_{uni[\overline{1}10]}=K_{uni[010]}=0$). Typically DW nucleation
and propagation dominates the magnetization reversal process, i.e.
$\varepsilon$ is much smaller than the crystalline anisotropy.
That is why it can be assumed that the magnetostatic energy minima
remain to a good approximation along the biaxial easy axes during
the double-step switching process. The energy difference between
stable magnetization states is thus given by the respective
difference in Zeeman energy (Eq. \ref{eqn:E}). When the energy
gained through a $90^\circ$ magnetization reorientation is larger
than $\varepsilon_{90^\circ}$, the nucleation and propagation
energy of a $90^\circ$-DW, a thermally activated switching event
becomes possible. This, on the timescale of our measurement,
results in an immediate switching event. For example, to calculate
the field required for the magnetization to jump from 0$^\circ$ to
90$^\circ$, the difference in the Zeeman terms is equated with
$\varepsilon$
\begin{center}
{\parbox{15cm}{
\begin{equation}
\label{eqn:DeltaE=eps} \Delta E_{0^\circ\rightarrow90^\circ}
=-MH[\cos(0^\circ-\varphi)-\cos(90^\circ-\varphi)]
=\varepsilon_{90^\circ}>0.
\end{equation}
}}
\end{center}
Reorganizing gives the switching field $H_{c}$ as a function of
$\varphi$.
\begin{center}
{\parbox{15cm}{
\begin{equation}
\label{eqn:Hc_biaxial}
H_{c}=\frac{-\varepsilon_{90^\circ}}{M[\cos\varphi-\sin\varphi]}
\end{equation}
}}
\end{center}
This equation is the same for other pairs of angles, except for
the signs in front of the sine and cosine functions in the
denominator, in the following marked with $\pm$. The switching
field equation above describes straight lines if plotted in a
polar coordinate system using H as radial and $\varphi$ as angular
coordinate. The polar plot in Fig.~\ref{fig:polarLinesBiaxial}
shows the resulting characteristic square pattern\cite{Cowb95}.
We express the switching field positions in this plot (thick
lines) in cartesian coordinates using $x=H_{c} \cos\varphi$ and $
y=H_{c} \sin\varphi$ to get a better feeling for the switching
field behaviour and to extract important parameters.
\begin{center}
{\parbox{15cm}{
\begin{equation}
\label{eqn:multiline}
\begin{tabular}{ r c l }
  \(H_c\cdot M[\pm\cos\varphi \pm\sin\varphi]\) & \(=\) & \(-\varepsilon_{90^\circ}\) \\
  \(M [\pm x\pm y]\) & \(=\) & \(-\varepsilon_{90^\circ}\) \\
  \(y\) & \(=\) & \(\pm x \pm \frac{\varepsilon_{90^\circ}}{M}\)
\end{tabular}
\end{equation}
}}
\end{center}
The characteristic polar-plot-pattern for a biaxial material is
thus a square with diagonals along the easy axes (the coordinate
axes in Fig.~\ref{fig:polarLinesBiaxial}). The first switching
field (thick blue lines) is largest along the easy axes, where
H$_{c1}=\varepsilon/M$. The DW nucleation/propagation energy can
be extracted from the diagonal of the square, whose length is
equal to $\frac{2\varepsilon_{90^\circ}}{M}$. All switching field
lines in Fig.~\ref{fig:polarLinesBiaxial} have an angle of
45$^\circ$ to the coordinate axes. The dashed lines represent the
hard magnetic axes. Arrows illustrate the direction of the
magnetization and their color the corresponding resistance state
of the respective section in an AMR measurement with current along
one of the easy axes.

Neglecting coherent rotation is typically a good model for the
first switching fields $H_{c1}$, whereas $H_{c2}$ is influenced by
magnetization rotation especially close to the hard axes. Pairs of
parallel lines in Fig.~\ref{fig:polarLinesBiaxial} do not extend
to infinity in practice, they move closer to the hard axes (see
the figures and the discussion in section~\ref{ssec:FPGallery}).
The magnetic field needed to force the magnetization parallel to
the external field in the hard axis direction is called the
anisotropy field $H_a$. It is a measure of the anisotropy strength
and can be calculated from Eq.~\ref{eqn:E} using the definition of
the anisotropy field: $H_a$ is the strength of a field along the
hard axis (here 45$^\circ$) needed to suppress the local minima
along the easy axes.
\begin{center}
{\parbox{15cm}{
\begin{equation}
\label{eqn:Ha_biaxial} H_{a}=\frac{2K_{cryst}}{M}
\end{equation}
}}
\end{center}

\subsection{The Anisotropy Fingerprint
Technique}\label{sec:fingerprints}

Traditionally the magnetic anisotropy is investigated by direct
measurement of the projection of the magnetization onto
characteristic directions using SQUID or VSM. The advent of vector
field magnets has recently opened up possibilities for acquiring a
detailed mapping of the anisotropy. We introduced such a method,
which builds on the above discussed angular dependence of the
magnetization switching fields, in Ref. \cite{fingerprint} and
expand upon it here. It is based on summarizing the results of
transport measurements into color coded resistance polar plots
(RPP) which act as fingerprints for the anisotropy of a given
structure. Not only is this method faster than the traditional
alternatives, but it is also more sensitive to certain secondary
components of the anisotropy, in particular those with easy axes
collinear to the primary biaxial anisotropy
component\cite{Cowb95}. The technique thus often reveals the
existence of components which would be missed using standard
techniques. Moreover, the technique can be applied to study the
anisotropy of layers by using macroscopic transport structures, or
applied directly to device elements. It can reveal effects of
processing or the influence of small strain fields due to, for
example, contacting.

\begin{figure}[t]
\centering
\includegraphics[width=14cm]{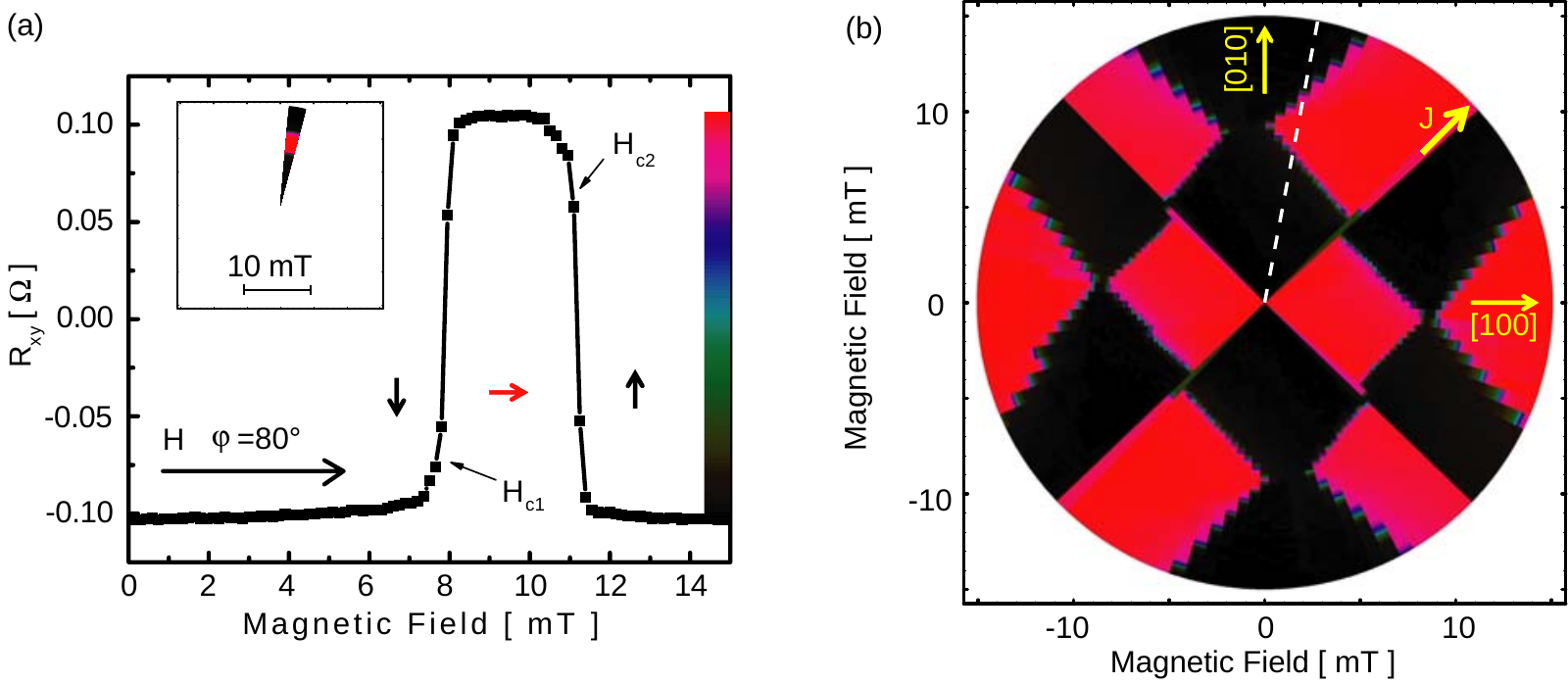}
\caption{\label{Fig:FP-meas} \textit{a) Planar Hall Effect
measurement along $\varphi=80^{\circ}$ with marked first and
second switching field, color scale and the corresponding section
of a color coded resistance polar plot(inset). b) Resistance polar
plot from a full set of Planar Hall measurements along every
$3^\circ$. The $80^\circ$-section corresponding to (a) is marked
by a dashed line.}} \vspace*{-0.2cm}
\end{figure}

In the present case the planar Hall effect is used to monitor the
magnetization behaviour in a standard Hall bar oriented along the
$\langle110\rangle$ crystal direction. Fig.~\ref{Fig:FP-meas}a
shows a planar Hall scan along $\varphi=80^{\circ}$. After
magnetizing the sample at -300 mT along $80^{\circ}$, the field is
brought down to zero. The figure shows the typical double-step
switching behaviour as discussed previously in connection with
Fig.~\ref{fig:AMRsim}b. The arrows indicate the magnetization
direction in the respective field regions with respect to the
crystal directions given in Fig.~\ref{Fig:FP-meas}b. Abrupt jumps
in resistance mark the first and second switching field $H_{c1}$
and $H_{c2}$. The normalized resistance value is color coded
according to the scale in Fig.~\ref{Fig:FP-meas}a. It is plotted
in a polar coordinate system along the magnetic field direction
$\varphi$ and with the magnetic field as radial scale. The inset
of Fig.~\ref{Fig:FP-meas}a shows the polar plot section
corresponding to the $80^{\circ}$-scan in this figure. Such Planar
Hall scans are recorded along many different in-plane field
directions and summarized in the resistance polar plot (RPP) in
Fig.~\ref{Fig:FP-meas}b. The $80^\circ$-segment is marked by a
dotted white line.

\begin{figure}[t]
\centering
\includegraphics[width=16cm]{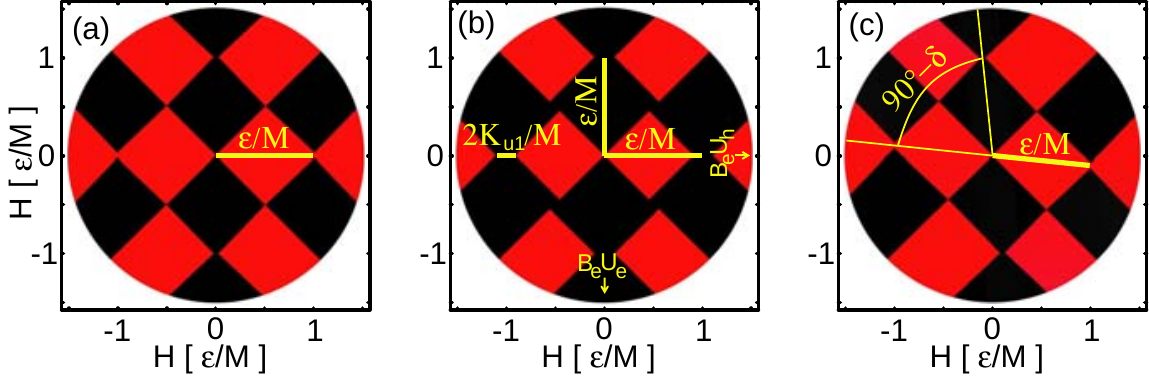}
\caption{\label{fig:FPs} \textit{Calculated resistance polar plots
for a biaxial material with easy axes along the [100](0$^\circ$)
and [010](90$^\circ$) crystal directions(a) and the same material
with an additional uniaxial anisotropy along [010](b) or $[1
\overline{1}0]$(c). Color scale of the resistance as in
Fig.~\ref{Fig:FP-meas}. $\varepsilon$ denotes the $90^\circ$-DW
nucleation/propagation energy.} }
\end{figure}

 We can now compare
the observed switching field pattern in Fig.~\ref{Fig:FP-meas}b
with the calculated shape in Fig.~\ref{fig:polarLinesBiaxial}.
While a cursory examination suggests a similar H$_{c1}$-pattern, a
more detailed comparison reveals significant differences:
Focussing on the innermost switching event, the pattern is indeed
strongly square-like, confirming that the (Ga,Mn)As has a mainly
biaxial magnetic anisotropy at 4 K. The diagonals of this
square-like H$_{c1}$-pattern are close to the [100] and the [010]
crystal direction, the easy axes of the biaxial anisotropy term.
However, the inner region is elongated (a rectangle and not a
square) - the signature of an additional uniaxial anisotropy term
with an easy axis bisecting the biaxial easy axes
(Fig.~\ref{fig:FPs}c), as will be discussed in
section~\ref{ssec:FP:uni110}. Additionally we observe a
discontinuity in the middle of the rectangle sides and dark "open"
corners close to the [010] direction. This is characteristic of a
uniaxial magnetic anisotropy term collinear with one of the
biaxial easy axes (Fig.~\ref{fig:FPs}b) and will be discussed in
detail in section~\ref{ssec:FP:uni010}.

These qualitative changes in the behaviour of H$_{c1}$ are key
signatures of the different anisotropy terms of the (Ga,Mn)As
layer. A set of high resolution transport measurements compiled
into a color coded resistance polar plot thus constitutes a
fingerprint of the symmetry components of the anisotropy. It
allows for the qualitative and quantitative determination of the
different anisotropy terms. It can prove their existence and
visualize their respective effects on the magnetization reversal.

\subsection{Signature of a $\langle010\rangle$ Uniaxial Term}\label{ssec:FP:uni010}

The fingerprint of a magnetically biaxial material in
Fig.~\ref{fig:FPs}a is equivalent to the switching field pattern
in Fig.~\ref{fig:polarLinesBiaxial}. If an additional small
uniaxial anisotropy $K_{uni[010]}$ along one of the biaxial easy
axes (here along $90^\circ$) is present, the square pattern is
altered as shown in Fig.~\ref{fig:FPs}b. The four-fold symmetry is
broken, and the biaxial easy axes correspond to energy minima of
slightly different depth, because one of them is parallel (biaxial
easy, uniaxial easy; $B_eU_e$) and one perpendicular (biaxial
easy, uniaxial hard; $B_eU_h$) to the easy axis of the uniaxial
anisotropy component.

The angle dependent switching field can be derived as discussed
above following \cite{Cowb95}: Again it is assumed, that the
minima of the energy surface remain at their zero field angles
throughout the switching process. In the present case however, the
energy minimum along the [010] direction is favored. Its energy is
$\Delta E=K_{uni[010]}$ smaller compared with the [100] direction,
which results in

\begin{center}
{\parbox{15cm}{
\begin{equation}
\label{eqn:Hc:uni010}
\begin{tabular}{ r c l }
  \(H_{c}\) & \(=\) & \(\pm \frac{\varepsilon_{90^\circ}\pm K_{uni[010]}}{M[\cos\varphi\pm\sin\varphi]}\) \\
  & & \\
  \(y\) & \(=\) & \(\pm x \pm \frac{\varepsilon_{90^\circ}}{M}\pm \frac{K_{uni[010]}}{M}\)
\end{tabular}
\end{equation}
}}
\end{center}

Magnetization reorientations towards the easier biaxial easy axis
$B_eU_e$ occur now at lower magnetic fields compared to the pure
biaxial anisotropy; switches away from $B_eU_e$ at higher fields.
The signs in eq.~\ref{eqn:Hc:uni010} are chosen appropriately. As
a result, the $H_{c1}$- pattern changes as displayed in
Fig.~\ref{fig:FPs}b. Characteristic features are the steps along
the biaxial hard axes, for example along $45^\circ$, and the
typical "open corners" along the $B_eU_e$ axis. These open corners
(in black along $90^\circ$ in Fig.~\ref{fig:FPs}b) arise because a
$180^\circ$-magnetization reorientation through the nucleation of
a $180^\circ$-DW is energetically favored in a small angular
region around the $B_eU_e$ axis \cite{Cowb95}.

Since the isotropic magnetoresistance\cite{physicaE} of typical
samples is relatively small compared to the AMR, two magnetization
directions differing by $180^\circ$ are not distinguishable on the
scale considered here, and have nearly the same color in the RPP,
creating the characteristic "open corner". The strength of the
uniaxial anisotropy component can be determined from the
separation $\frac{ 2K_{u1} }{M}$ between $H_{c1}$ and $H_{c2}$
along the $B_eU_h$ axis.

\subsection{Signature of a $\langle \overline110 \rangle$ Uniaxial Term}\label{ssec:FP:uni110}

\begin{figure}[t]
\centering
\includegraphics[width=16cm]{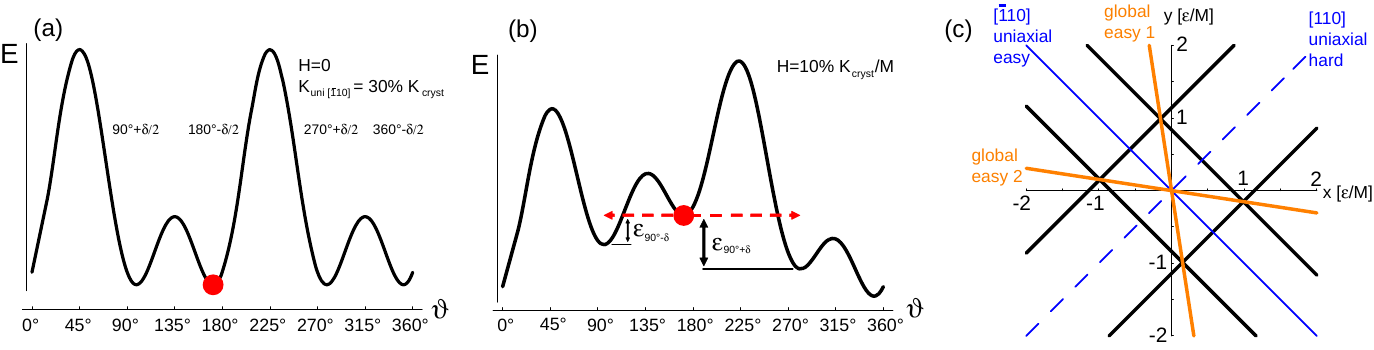}
\caption{\label{fig:110lines} \textit{a) A uniaxial
$[\overline{1}10]$ anisotropy component flattens the energy
surface (eq.~\ref{eqn:E}) and shifts the positions of the energy
minima. b) Energy landscape with magnetic field applied along the
$-\delta/2$-global easy axis direction. A clockwise and
counterclockwise jump of the magnetization (with the respective
$\varepsilon$) are equally possible. c) Switching field positions
in the polar plot (thick black lines), global easy axes (orange)
and easy and hard direction of the $[\overline{1}10]$ anisotropy
component (blue).} }
\end{figure}

In this section we describe the effects of a uniaxial anisotropy
term $K_{uni[\overline{1}10]}$ with its easy axis (along
135$^\circ$) bisecting the biaxial easy axes. This uniaxial
anisotropy component flattens the energy surface (eq.~\ref{eqn:E})
and shifts the positions of the energy minima by (see
Fig.~\ref{fig:110lines}a)
\begin{center}
{\parbox{15cm}{
\begin{equation}
\label{eqn:delta}
\frac{\delta}{2}=\frac{1}{2}\arcsin(\frac{K_{uni[\overline{1}10]}}{K_{cryst}})
\end{equation}
}}
\end{center}
towards the uniaxial easy axis\cite{Daboo}. All four minima have
the same energy value. To derive the switching fields we equate
the DW nucleation/propagation energy $\varepsilon$ with the
difference in Zeeman energy between the initial and final
magnetization angle in the respective switching event. As
illustrated in Fig.~\ref{fig:110lines}a, the magnetization
direction can change by 90$^\circ+\delta$ or 90$^\circ-\delta$
depending on whether the magnetization rotates clockwise or
counterclockwise. Following \cite{Roukes} we use different DW
nucleation/propagation energies $\varepsilon_{90^\circ+\delta}$
and $\varepsilon_{90^\circ-\delta}$ respectively. The switching
field positions in the polar plot given in cartesian coordinates
are
\begin{center}
{\parbox{15cm}{
\begin{equation}
\label{eqn:Hc:110}
\begin{tabular}{ r c l }
  \(y_{90^\circ+\delta}\) & \(=\) & \( x \pm \frac{\varepsilon_{90^\circ+\delta}}{M \sqrt{2}[\cos(45^\circ-\delta/2)]}\) \\
  & & \\
  \(y_{90^\circ-\delta}\) & \(=\) & \( -x \pm \frac{\varepsilon_{90^\circ-\delta}}{M \sqrt{2}[\cos(45^\circ+\delta/2)]}\) \\
\end{tabular}
\end{equation}
}}
\end{center}
 Equation~\ref{eqn:Hc:110} describes two parallel sets of lines, as shown in
 Fig.~\ref{fig:110lines}c (thick black lines), whose distance from the origin is determined by
 the respective $\varepsilon$. MOKE experiments on epitaxial iron films grown
 on GaAs (with similar anisotropy terms as (Ga,Mn)As) confirm that as expected the sense of the magnetization rotation
 changes when crossing a global easy axis \cite{Daboo}. The two line sets of eq.~\ref{eqn:Hc:110}
 represent the clockwise and counterclockwise sense of magnetization rotation. If the field is
 applied along a global easy axes (minima of
 Fig.~\ref{fig:110lines}a) both rotation directions are energetically
 equivalent. Consequently the lines must intersect along global easy axes directions.
 Fig.~\ref{fig:110lines}b shows the energy landscape of Fig.~\ref{fig:110lines}a
 when a magnetic field is applied along the $-\delta/2$-global easy axis
 direction. For both rotation directions, the Zeeman term at the first switching field $H_{c1}$ is
 equal to the respective $\varepsilon$.
 We can thus calculate the dependence of $\varepsilon$ on the angle $\Delta
 \vartheta$ between initial and final magnetization direction:
\begin{center}
{\parbox{15cm}{
\begin{equation}
\label{eqn:eps}
\begin{tabular}{ r c l }
  \(\varepsilon_{90^\circ \pm \delta}\) & \(=\) & \( H_{c1}M (1-\cos(90^\circ \pm \delta)) \) \\
  & & \\
  \(\varepsilon_{\Delta \vartheta}\) & \(=\) & \( \varepsilon_{90^\circ}(1- \cos(\Delta \vartheta)) \) \\
\end{tabular}
\end{equation}
}}
\end{center}

which is intuitively reasonable. At the same time we find, that
$H_{c1}$ along a global easy axis is $\varepsilon_{90^\circ}/M$.
Note that this careful treatment of $\varepsilon$ is necessary,
the simplified model of a constant $\varepsilon$ independent of
the DW angle $\Delta\vartheta$, would lead to the incorrect
conclusion, that the rectangle in the polar plot would have its
long axis perpendicular to the uniaxial easy direction.

A summary of the above is shown in Fig.~\ref{fig:FPs}c. The
characteristic pattern of a mainly biaxial anisotropy with a
bisecting uniaxial anisotropy component is rectangular. The
diagonals of the rectangle are the "global easy axes", their
length is 2$\varepsilon_{90^\circ}/M$. The angle between the
global easy axes gives the relative strength of the two anisotropy
components (using eq.~\ref{eqn:delta}). The easy axis of the
uniaxial term is along the median line of the longer edge, for
example along 135$^\circ$ in Fig.~\ref{fig:FPs}c.

\begin{figure}[tbp]
\centering
\includegraphics[width=16cm]{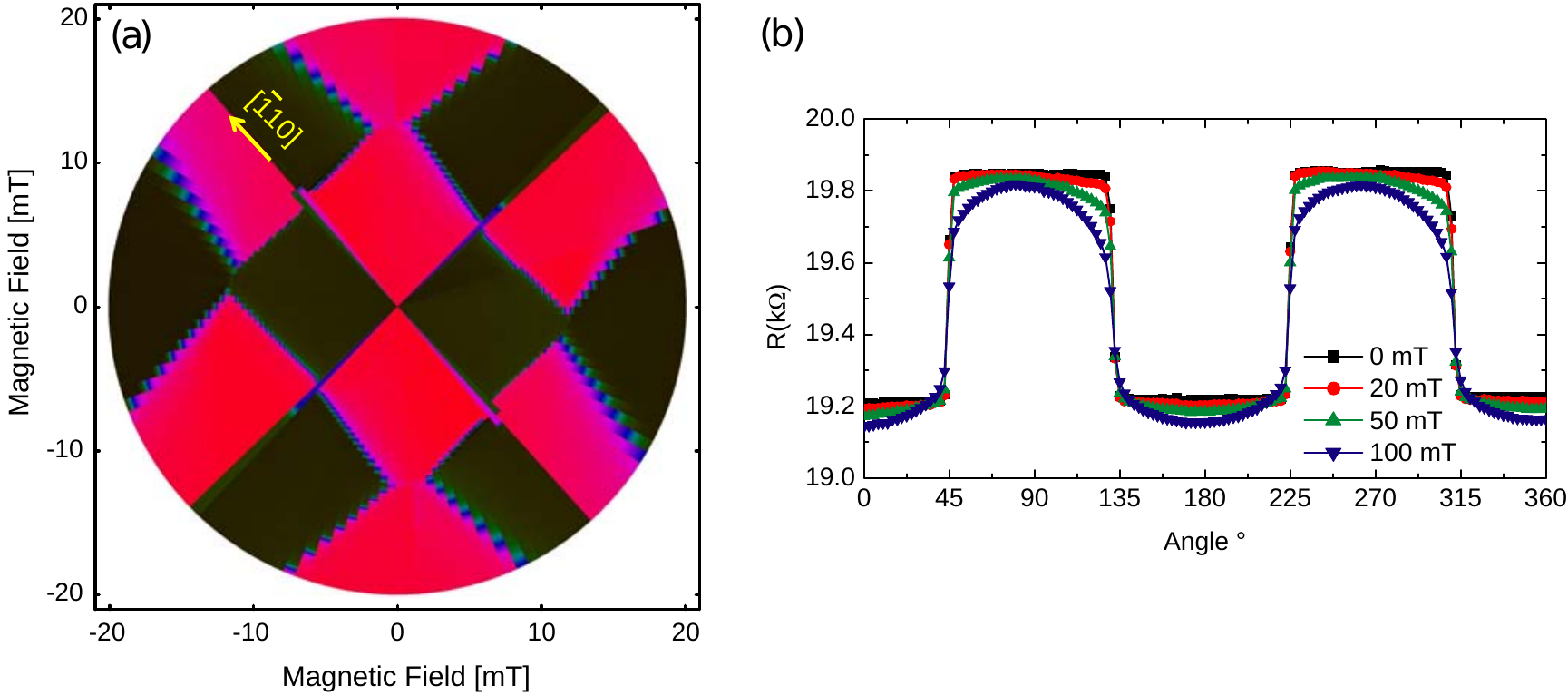}
\caption{\label{fig:110phiscan} \textit{a) Fingerprint of a
typical 20 nm thick (Ga,Mn)As Hall bar and b) angle-dependent
 longitudinal resistance at different fields after magnetizing along $\varphi$.}}
\end{figure}

The presence and sign of the $\langle\overline{1}10\rangle$
anisotropy term can be verified with the help of AMR or PHE
measurements at magnetic fields of medium amplitude. For
comparison, longitudinal resistance measurements on a Hall bar
sample oriented along a (Ga,Mn)As easy axis (0$^\circ$) are
converted into the RPP displayed in Fig.~\ref{fig:110phiscan}a.
This fingerprint shows an overall biaxial anisotropy with easy
axes close to 0$^\circ$ and 90$^\circ$. The central pattern is
elongated along 135$^\circ$, suggesting that a uniaxial anisotropy
component with easy axis along this direction (the
$[\overline{1}10]$ crystal direction) is present.

This is confirmed by the measurements in
Fig.~\ref{fig:110phiscan}b. Here the Hall bar sample is first
magnetized in a high magnetic field of 300 mT along an angle
$\varphi$. The longitudinal resistance is then measured, while the
field is slowly stepped down to zero. Fig.~\ref{fig:110phiscan}b
shows the resistance values at 100 mT, 50 mT, 20 mT and 0 mT as a
function of the field angle. For the interpretation of these
curves, imagine for example an energy landscape as shown in
Fig.~\ref{fig:110lines}, where the strength of the
$[\overline{1}10]$ uniaxial anisotropy term is exaggerated. This
term describes the width and the height of the "hills" in the
energy surface. The "hill" in the uniaxial easy axis direction
(here 135$^\circ$) is lower than the energy barrier perpendicular
to this direction, which is steeper and coincides with the hard
magnetic axis of the $[\overline{1}10]$ uniaxial term. At zero
field the magnetization is aligned with one of the biaxial easy
axes(black curve in Fig.~\ref{fig:110phiscan}b). The steps in this
curve mark the peak positions of the "hills" in the energy
landscape - the biaxial hard axes. At medium fields (e.g. 50 mT in
Fig.~\ref{fig:110phiscan}b), the magnetization is rotated away
from the global easy axes, causing deviations from the step-like
behaviour at zero field. These deviations occur at smaller field
values along the uniaxial easy direction $[\overline{1}10]$
compared with the uniaxial hard axis [110]. The direction (meaning
the sign of $K_{uni[\overline{1}10]}$) of the $[\overline{1}10]$
uniaxial anisotropy is thus the same as in
Fig.~\ref{fig:110phiscan}a: the abrupt resistance change at
45$^\circ$ marks the hard and the smoother behaviour at
135$^\circ$ the easy uniaxial axis direction.

\subsection{(Ga,Mn)As at 4
K - Typical Fingerprints }\label{ssec:FPGallery}

\begin{figure}[tbp]
\centering
\includegraphics[width=7cm]{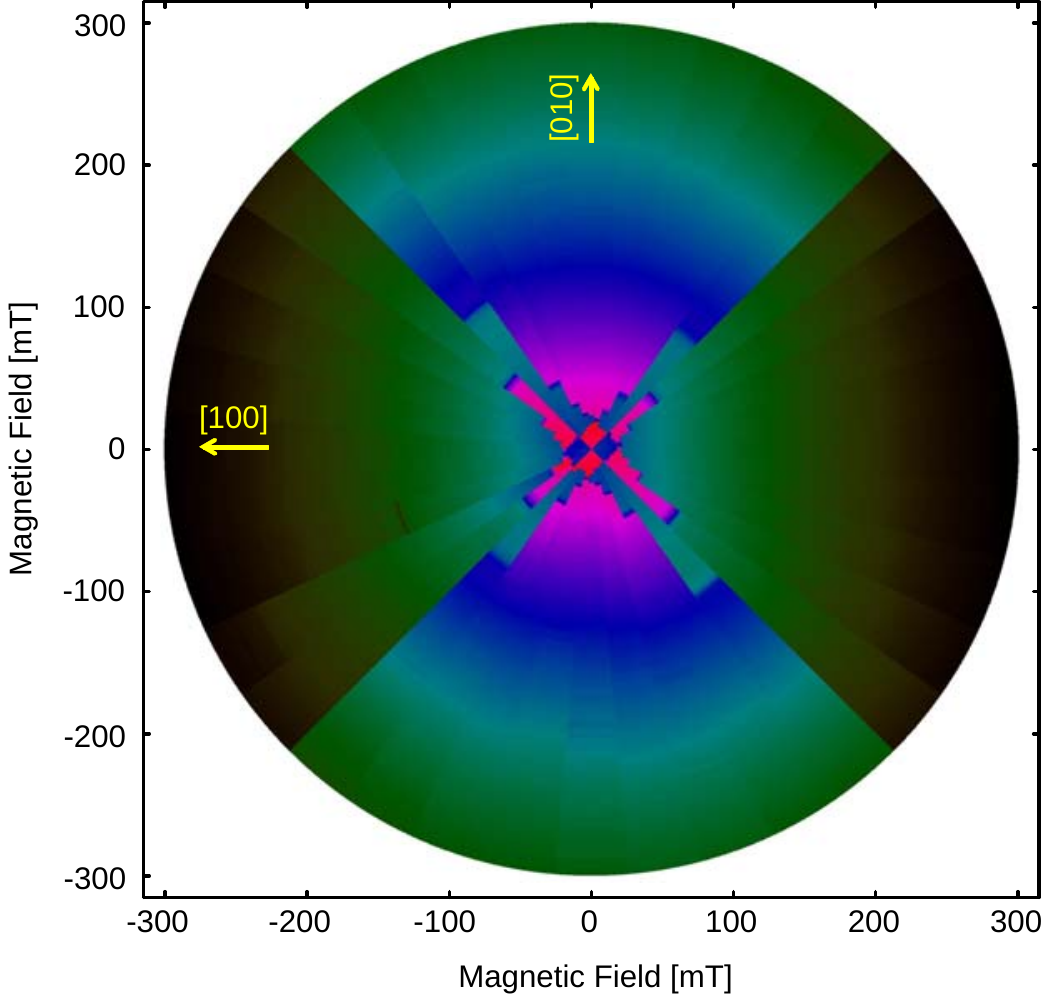}
\caption{\label{fig:FP-wuebadd} \textit{ Typical AMR fingerprint
measurement of a 100 nm thick (Ga,Mn)As Hall bar. The current
direction is
along 0$^\circ$.} }
\end{figure}

In the above sections we describe a method which is sensitive
enough to detect both, the $[\overline{1}10]$ and the [010]
uniaxial anisotropy term. Here we apply the method to our typical
(Ga,Mn)As layers and find that all three anisotropy components,
the biaxial and the two uniaxial ones, are present in every
sample. Various fingerprints show the typical variation of the
relative anisotropy terms and the characteristics at low and high
fields.

\begin{figure}[tbp]
\centering 
\includegraphics[width=14cm]{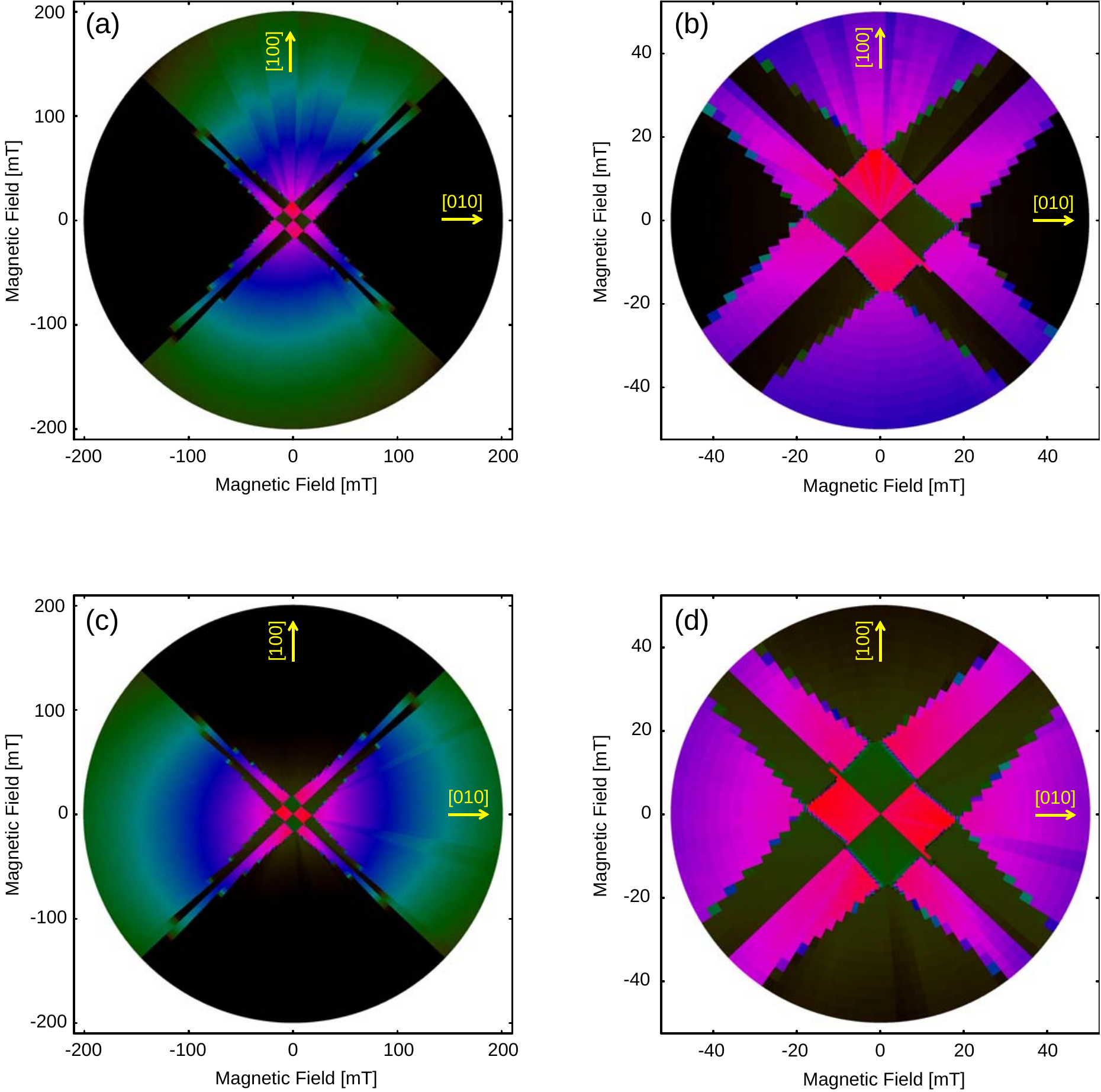}
\caption{\label{fig:FP-wueb} \textit{High angular resolution
fingerprint measurements (a,b) and close ups of the central region
(c,d) for two Hall bars made of the same 70 nm thick material but
oriented along orthogonal crystal directions. The current flows
along
0$^\circ$ in a and b and along 90$^\circ$ in c and d.} } 
\end{figure}

The fingerprints in Figs.~\ref{fig:110phiscan}a,
\ref{fig:FP-wuebadd} and \ref{fig:FP-wueb} were compiled from
longitudinal AMR measurements on typical (Ga,Mn)As layers of
different thickness. All these plots including
Fig.~\ref{Fig:FP-meas}b show the same general pattern, resembling
the four-fold switching field pattern in Fig.~\ref{fig:FPs}. The
main anisotropy component in all these layers at 4~K is thus biaxial
with easy axes along the [100] and [010] direction. The strength of
this biaxial term is measured by taking separate high resolution AMR
curves along the hard magnetic axes and concluding the anisotropy
constant from the anisotropy fields. Typical 2K/M values are of the
order of 100~mT...200~mT, as can be seen e.g. in the high field
fingerprints in Figs.~\ref{fig:FP-wuebadd} and ~\ref{fig:FP-wueb}a
and c, where in the sections along the hard axes the magnetization
is aligned with the external field at these field values.

\begin{figure}[tbp]
\includegraphics[width=14cm]{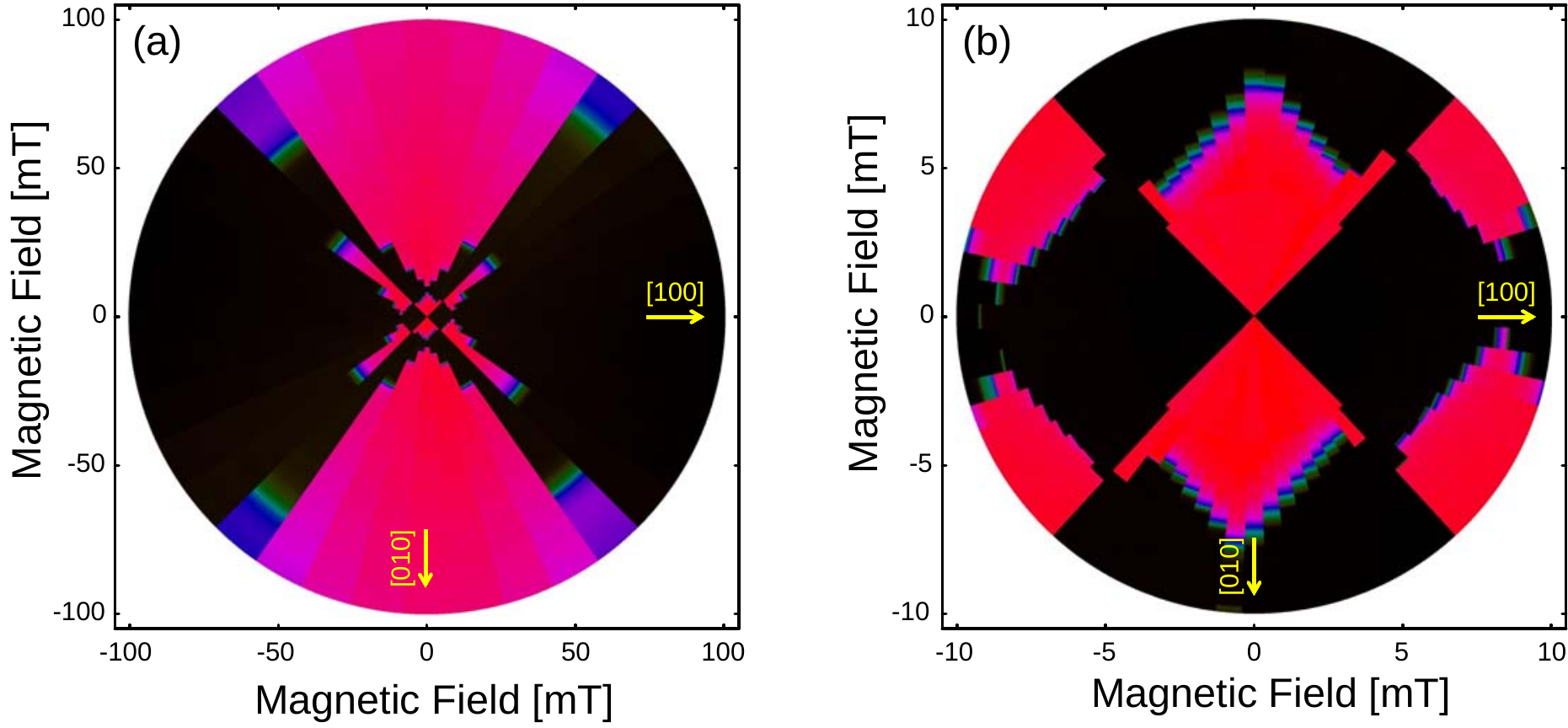}
\caption{\label{fig:FP-wueb2} \textit{Fingerprint measurement and
high resolution RPP at low magnetic field for a 70~nm thick
(Ga,Mn)As layer with strongly visible [010] uniaxial anisotropy
component but a relatively small $[\overline{1}10]$ term. } } 
\end{figure}

The anisotropy components and $\varepsilon$ differ of course from
wafer to wafer. The general pattern on the other hand is very
similar. All of the RPP show clearly an elongation of the
$H_{c1}$-pattern into a rectangle, the signature of the
$[\overline{1}10]$ anisotropy component. Steps along the hard axes
and the typical open corner are also always present, the typical
feature of the [010] anisotropy component. Both uniaxial
anisotropy components are thus clearly present in all investigated
samples.

The crystal directions are indicated in all fingerprints in
yellow. We find, that the elongation of the central pattern, and
thus the easy axis of the $[\overline{1}10]$ uniaxial component,
points along the $[\overline{1}10]$ crystal direction in all shown
typical transport samples. At this point, we would like to note,
that the sign of the $[\overline{1}10]$ uniaxial component, i.e.
whether the easy axis points along $[\overline{1}10]$ or [110],
depends on carrier concentration and Mn doping as shown by
\cite{SawickiEdmonds}. The elongation could thus also be along the
[110] direction depending on growth conditions and a possible
annealing treatment. In the as grown samples investigated here at
4~K, the typical strength of the $[\overline{1}10]$ uniaxial
anisotropy is of the order of 10\% of the biaxial anisotropy
constant. As examples of the range of values typical for this
ratio varies we note a value of 10\% from Fig.~\ref{fig:FP-wueb2},
15\% from Fig.~\ref{Fig:FP-meas}b and 20\% from
Fig.~\ref{fig:FP-wueb}.

The open corners of the $H_{c1}$-pattern indicate the direction of
the easy axis of the [010] uniaxial term. This easy axis direction
is sample dependent and can be along either of the biaxial easy
axes of the sample. In the shown polar plots we see this easy axis
along [010] in Fig.~\ref{Fig:FP-meas}b and along $[100]$ in
Figs.~\ref{fig:110phiscan},~\ref{fig:FP-wuebadd},~\ref{fig:FP-wueb}
and~\ref{fig:FP-wueb2}. Also the strength of the [010] term is
sample dependent. It can dominate the low field switching
behaviour as for example in Fig.~\ref{fig:FP-wueb2} or be barely
visible as in Fig.~\ref{fig:110phiscan}a. In any case, the
strength of this anisotropy component is extremely small compared
to the main biaxial anisotropy. Even in Fig.~\ref{fig:FP-wueb2},
where the presence of the [010] uniaxial component has a strong
influence on the magnetization behaviour at low fields, its
anisotropy field $2K_{uni[010]}/M$ is only 1.6~mT, only 1\% of the
typical biaxial anisotropy constant.

Figures~\ref{fig:FP-wueb}a and c show similar AMR fingerprints on
two Hall bars made from the same wafer, but oriented along
orthogonal crystal directions. Panels c and d show a close up of
the central region. Both fingerprints show virtually the same
switching pattern with inverted colors because of the orthogonal
current directions. This shows the high homogeneity of the wafer
and the robustness of the method. Even on several cool downs, we
see virtually the same switching pattern (not shown), although the
resistance of the sample changes slightly upon recooling. Note
that shape anisotropy in these structures is negligible compared
with the crystalline magnetic anisotropy contribution as discussed
in Ref.~\cite{nanobars}. It is too small to play any significant
role in these measurements.

We have neglected coherent rotation to derive simple formulas for
the switching fields in the polar plots. This is typically a good
model for the first switching fields $H_{c1}$, whereas $H_{c2}$ is
influenced by magnetization rotation especially along the hard
axes, as mentioned above. The effects of coherent rotation are of
course taken into account in the numerical modelling that is based
on the energy equation \ref{eqn:E}.

As discussed previously, the extent of the $H_{c1}$-pattern is
determined by $\varepsilon$, while the extent of the
$H_{c2}$-features is mainly given by the biaxial anisotropy
constant through the anisotropy field (see
eq.~\ref{eqn:Ha_biaxial}). If the ratio of biaxial anisotropy to
$\varepsilon$ is very large, the central pattern is well described
by DW-nucleation/propagation-related switching events alone. In
the higher field region, the minima of the energy surface move
considerably and the $H_{c2}$-switching events approach the hard
axes directions.

The typical situation is for example seen in
Fig.~\ref{fig:FP-wuebadd}, where AMR curves along every 10$^\circ$
were taken on a Hall bar along 0$^\circ$. The central region shows
a rectangular pattern (signature of the $[\overline{1}10]$
uniaxial term) with open corners and steps along the hard axes
(signature of the [010] uniaxial term). There is almost no
coherent rotation at these low fields. Magnetization
reorientations occur through DW nucleation and propagation as seen
from the abrupt color changes (between red and blue). The second
switching fields along the hard axes (e.g. along 45$^\circ$ at
50~mT) are marked by smooth color transitions proving that
coherent rotation is at play. Smooth color transitions at even
higher fields (green to black around 0$^\circ$ and red to green
around 90$^\circ$) finally are caused by the isotropic MR effect
\cite{physicaE}.

The fingerprint in Fig.~\ref{fig:110phiscan} shows a slightly
different situation. The ratio of the anisotropy energy to
$\varepsilon$ cannot be treated as infinite. For this reason also
the $H_{c1}$-pattern shows a considerable influence of coherent
rotation. The sides of the rectangle are no longer parallel to
each other and the corners do not draw an angle of 90$^\circ$.
Still, the elongation is obvious and the difference between
switching events towards the two easy axes is observable.

In summary we have shown a variety of fingerprints of typical
(Ga,Mn)As transport layers at 4~K. The fingerprint method allowed
us to identify the simultaneous presence of the biaxial and two
uniaxial anisotropy components. Indeed all (Ga,Mn)As layers
investigated show both these uniaxial components, including layers
where the [010] component could not be identified in SQUID
measurements. As a rule of thumb, the typical relative strength of
the anisotropy terms is of the order of
$K_{cyst}:K_{uni[\overline{1}10]}:K_{uni[010]}\sim 100:10:1$.

\section{Uniaxial Magnetic Anisotropy}\label{sec:el-uniaxial}

This section deals with the magnetization behaviour of
magnetically uniaxial materials and how it manifests itself in
transport measurements. The latter with two specific applications
in mind:
\begin{itemize}
\item The (Ga,Mn)As magnetic anisotropy is strongly temperature
dependent with the $\langle 110 \rangle$ uniaxial anisotropy term
being dominant close to the Curie temperature
(section~\ref{ch:T-dep}). \item The fingerprint method can also be
used to characterize individual transport structures or even
device components. Uniaxial magnetic behaviour was recently
achieved by submicron patterning of (Ga,Mn)As and the
corresponding anisotropic strain relaxation \cite{nanobars}.
\end{itemize}

We again track the magnetization angle using AMR measurements and
finally discuss the color-coded RPP, the anisotropy fingerprint,
expected for a material with uniaxial magnetic anisotropy.

\begin{figure}[t]
\centering
\includegraphics[width=12cm]{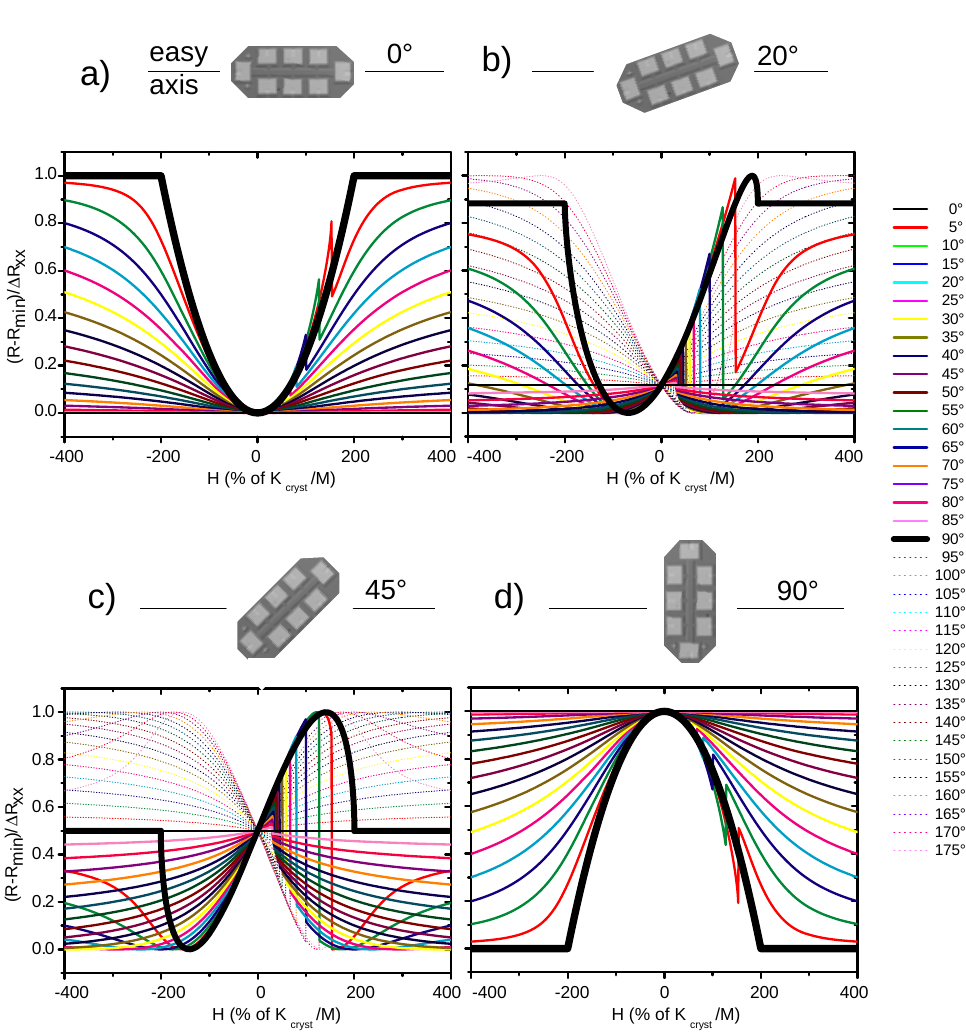}
\caption{\label{fig:unipanels} \textit{ Calculated anisotropic
magnetoresistance curves in a magnetically uniaxial material for
magnetic field sweeps along many in-plane
directions($0^{\circ}..85^{\circ}$ thin solid, $90^{\circ}$ thick,
$95^{\circ}..175^{\circ}$ dashed) for Hall bar orientations as in
the sketches with current along (a) $0^{\circ}$ , (b)
$20^{\circ}$, (c) $45^{\circ}$, and (d) $90^{\circ}$. All angles
with respect to the uniaxial easy axis. The field is swept from
left to right.} }
\end{figure}

Fig.~\ref{fig:unipanels} shows AMR curves calculated for a
magnetically uniaxial material using Eq.~\ref{eqn:E} with
$\varepsilon$ =30\% $K_{uni}$. The individual panels illustrate
how the current direction with respect to the easy axis modifies
the overall picture of a set of AMR curves. In all four panels a
single zero field resistance state can be identified,
corresponding to the easy axis magnetization orientation. The
resistance value is given by the angle between current and easy
axis through Eq.~\ref{eqn:Rxx}. If the external field is swept
along this easy axis direction ($0^\circ$, thin black line), the
magnetization is aligned with the easy axis throughout the whole
scan, yielding a horizontal line through the focal point at zero
field. The hard axis scan (thick line) reveals the anisotropy
field; (the same as in the biaxial case, Eq.~\ref{eqn:Ha_biaxial})
\begin{center}
{\parbox{15cm}{
\begin{equation}
\label{eqn:Ha_uniaxial} H_{a}=\frac{2K_{uni}}{M}
\end{equation}
}}
\end{center}
the external magnetic field perpendicular to the easy axis, where
the magnetization starts to deviate from the field direction. The
magnetization rotation in panels (a) and (d) yields a parabolic
dependence of the resistance on the field amplitude
\cite{NatureParabola}. In all other MR scans the magnetization
relaxes to the closest easy axis direction while the field is
decreased from high negative values, reaching the focal point at
zero field. After zero, the magnetization direction reverses by
circa $180^\circ$ through DW nucleation and propagation, which is
visible as abrupt resistance changes in Fig.~\ref{fig:unipanels},
for example the spikes around 100\% K$_{uni}$ in panel (a). A back
sweep results in a hysteretically symmetric curve with the
switching events at negative fields (not shown).

\begin{figure}[t]
\centering
\includegraphics[width=16cm]{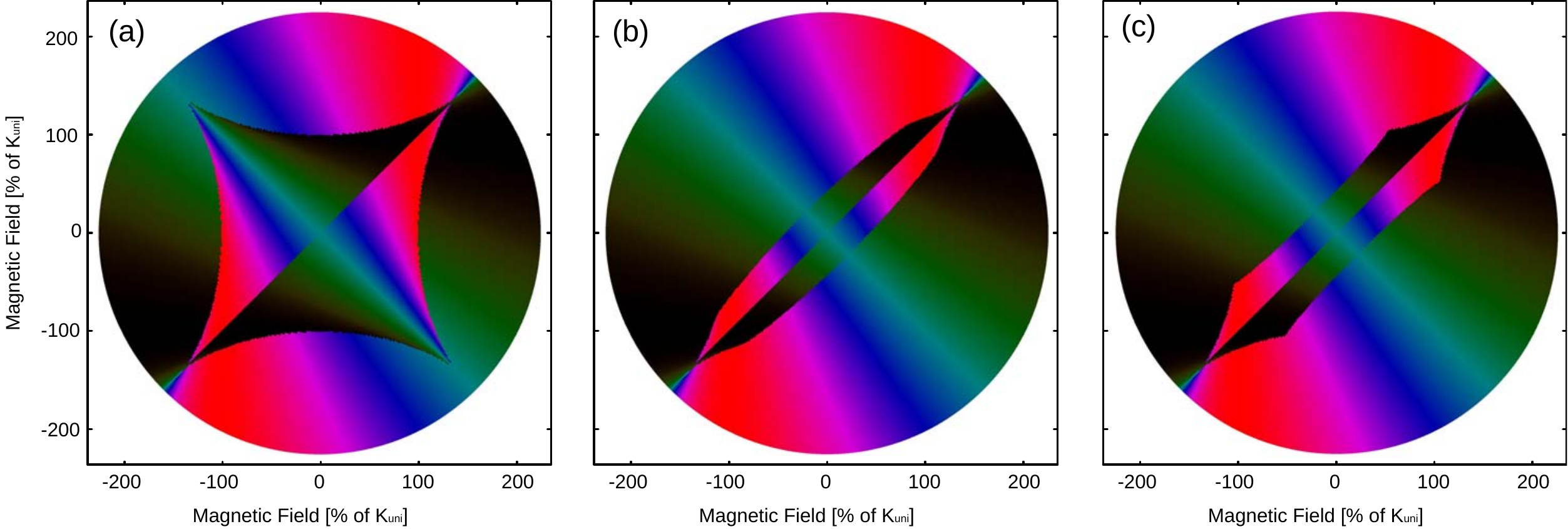}
\caption{\label{fig:FP-uniaxial} \textit{ Calculated AMR
fingerprints of a magnetically uniaxial material with easy axis
along $135^{\circ}$ and current along $0^{\circ}$. Magnetization
reversal through (a) coherent rotation only (Stoner-astroid) and
(b) DW nucleation and propagation with $\varepsilon$ according to
Eq.~\ref{eqn:eps} (c) simplified model assuming a constant
$\varepsilon=2\varepsilon_{90^\circ}$. ($\varepsilon_{90^\circ}$
=30\% $K_{uni}$)} }
\end{figure}

Fig.~\ref{fig:FP-uniaxial} shows the results of similar
calculations with high angular resolution plotted in RPP fashion.
Here the easy axis is oriented along $135^\circ$ and the current
flow along $0^\circ$. The colors are a function of the the current
direction, for example dark color at high magnetic fields along
the current, while the switching event pattern is defined by the
magnetic properties alone.

If a structure is smaller than the single-domain limit
\cite{Brown_singledomain,Aharoni_singledomain} it is energetically
unfavourable to nucleate a DW. Instead the magnetization rotates
coherently (Stoner-Wohlfarth model ~\cite{StonerWohlf}).
Fig.~\ref{fig:FP-uniaxial}a shows the well known Stoner-Wohlfarth
astroid \cite{StonerWohlf,Schaefer} which describes the switching
positions of a uniaxial particle under coherent rotation. Its
extent in both the easy and the hard axis direction is given by
the anisotropy field.

Allowing for DW nucleation with $\varepsilon$ according to
eq.~\ref{eqn:eps} truncates the easy axis corners of the astroid
as shown in Fig.~\ref{fig:FP-uniaxial}b. The extent
$\varepsilon_{90^\circ}/M$ in the easy axis direction is a measure
for the DW nucleation/propagation energy. A field sweep along the
hard magnetic axis, is still fully described by Stoner-Wohlfarth
rotation and the extent in this direction is given by the
anisotropy field. The detailed shape of the switching field
pattern depends on the model used for the $\varepsilon$-dependence
on the DW angle. Fig.~\ref{fig:FP-uniaxial}c shows the RPP
calculated assuming a constant $\varepsilon_{\Delta\vartheta}=2
\varepsilon_{90^\circ}$ independent of the magnetization
directions of the domains separated by the DW. While the easy and
hard axis extent are the same as in Fig.~\ref{fig:FP-uniaxial}b,
the better correspondence of the shape of the features in (b) then
(c) to the experimental data is further evidence in support for
the above described description of the DW energies.

\section{Temperature Dependence of the (Ga,Mn)As
Anisotropy}\label{ch:T-dep}

\begin{figure}[p]
\centering
\includegraphics[width=14cm]{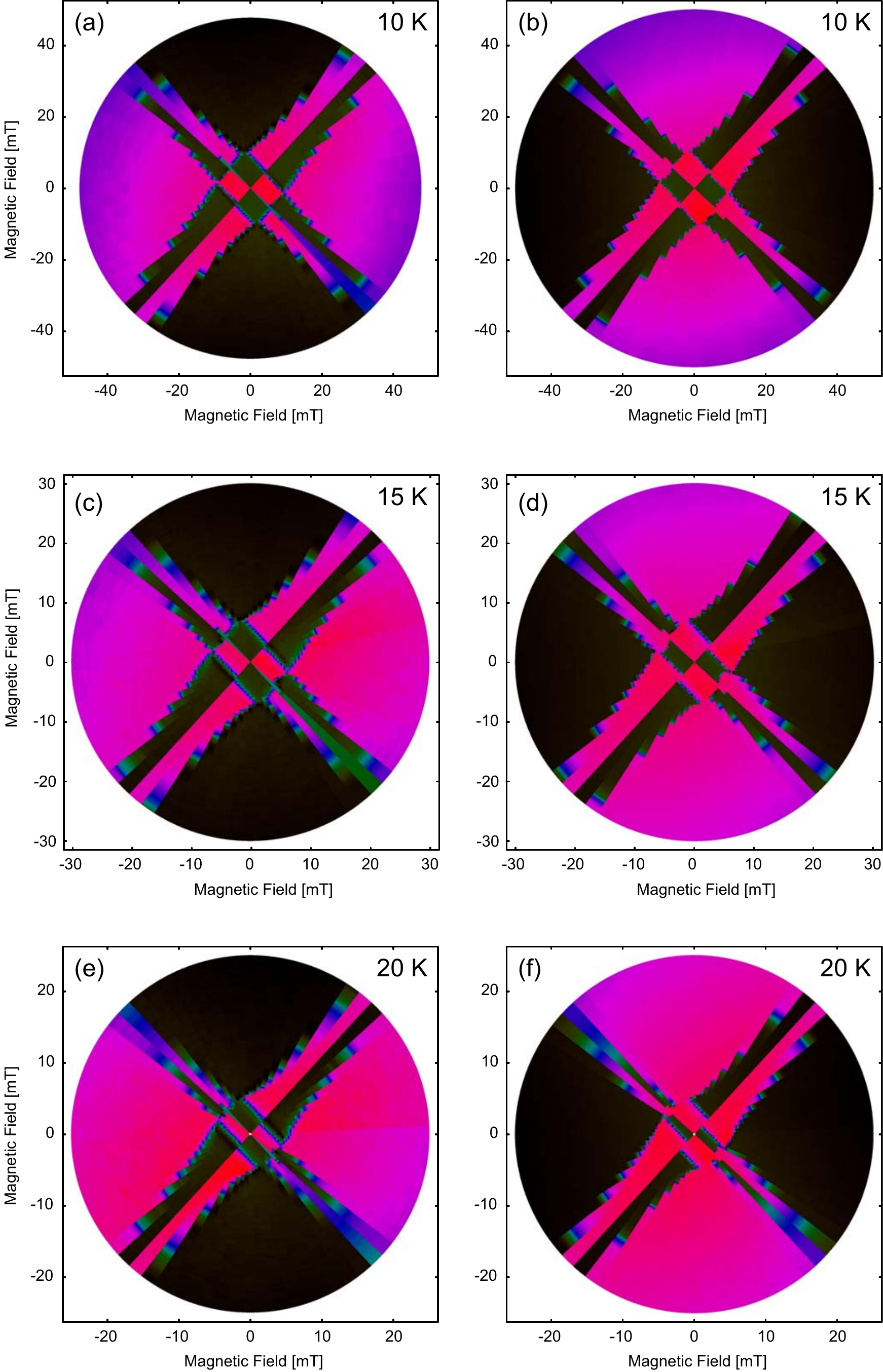}
\caption{\label{fig:FP-1} \textit{Temperature dependent AMR
fingerprint measurements of the sample in Fig.~\ref{fig:FP-wueb}a
and b (right column) with current along 0$^\circ$ and
Fig.~\ref{fig:FP-wueb}c and d (left column) with current along
90$^\circ$.}}
\end{figure}

\begin{figure}[p]
\centering
\includegraphics[width=14cm]{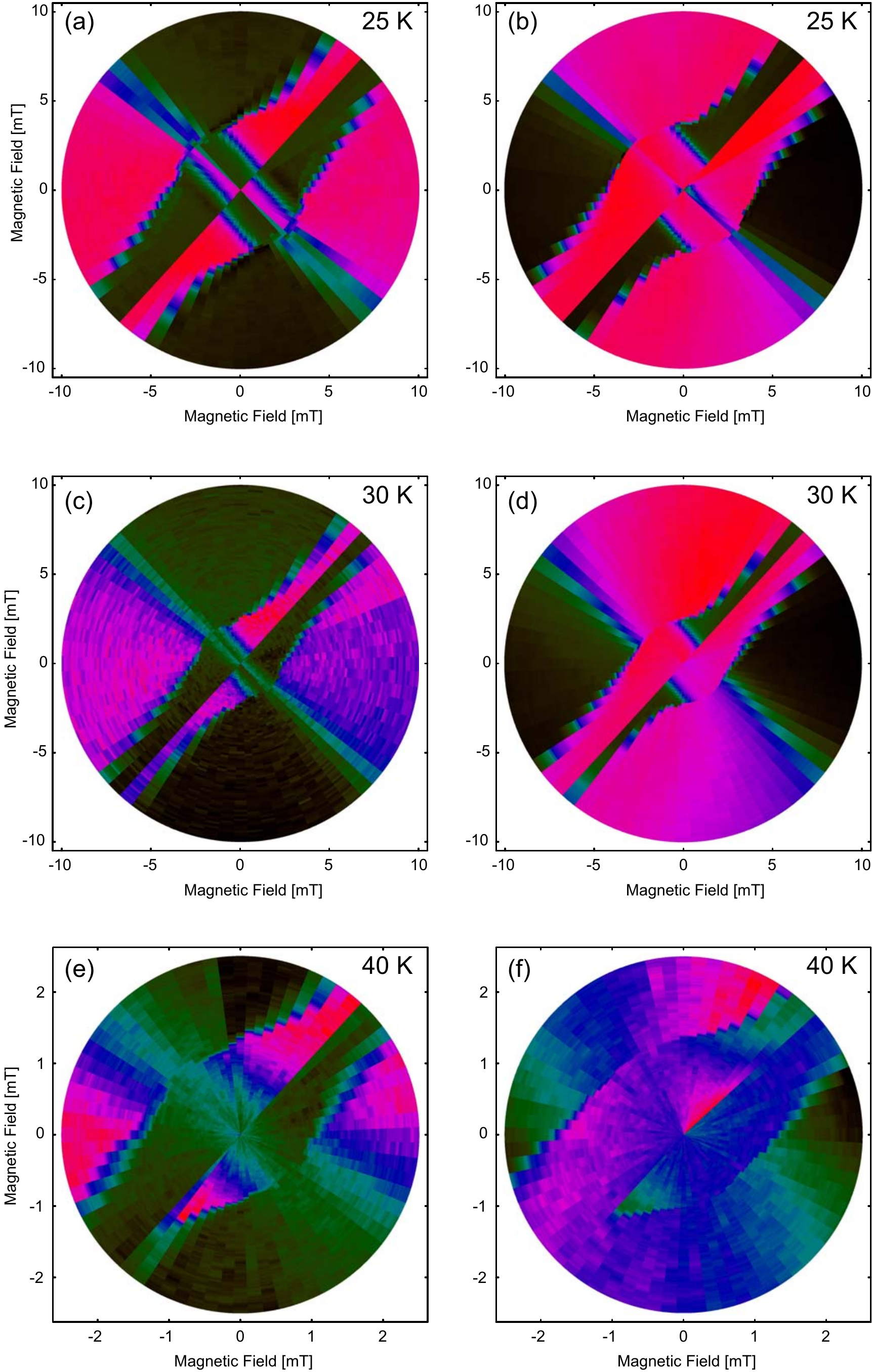}
\caption{\label{fig:FP-2} \textit{High temperature AMR fingerprint
measurements, continuation of Fig.~\ref{fig:FP-1}.} }
\end{figure}

The fingerprint method provides us with the opportunity to
investigate the temperature dependence of the magnetic anisotropy.
Figures~\ref{fig:FP-1} and~\ref{fig:FP-2} show AMR fingerprints at
various temperatures for the layer investigated in
Fig.~\ref{fig:FP-wueb} at 4.2 K. The left column shows results on
a Hall bar patterned along 90$^\circ$ (the [100] crystal
direction). In the right column the Hall bar is oriented along
0$^\circ$. The layer is, as typical, very homogeneous and the
switching patterns in the two columns are virtually identical at
all temperatures (except for a trivial inversion of the color
scales).

The mainly biaxial anisotropy is the origin of the nearly
four-fold symmetry in the low temperature fingerprints. The
uniaxial anisotropy term with easy axis along the
$[\overline{1}10]$ crystal direction takes over with increasing
temperature and becomes the dominant term close to T$_C$: already
the fingerprints at 30 K exhibit the typical 2-fold symmetry of a
uniaxial anisotropy. The short axis of the pattern marks the
uniaxial easy axis; the extended feature perpendicular to it the
magnetic hard axis (see Sec.~\ref{sec:el-uniaxial} for details).
The AMR amplitude and the switching fields, i.e. the size of the
fingerprint pattern, decrease significantly with temperature (note
the different magnetic field scales).

This is consistent with detailed SQUID studies
\cite{Mike05PRL,Mike05PRB}. There the anisotropy constants
$K_{cryst}$ and $K_{[\overline{1}10]}$ were extracted from hard
axis magnetization measurements vs magnetic field. The two terms
exhibited different temperature dependence. In particular it was
observed that the temperature dependence of the anisotropy
constants originates in their power-law dependence on the volume
magnetization $M$. While the uniaxial anisotropy constant is
proportional to the square of $M$, the biaxial term depends on
$M^4$. As a result, the biaxial anisotropy term, which dominates
the magnetic behaviour at 4 K, decreases much faster with
increasing temperature than the uniaxial term. This is the reason
why the magnetic anisotropy undergoes a transition from mainly
biaxial to mainly uniaxial when the temperature increases from 4 K
to T$_C$.

\begin{figure}[tb]
\centering
\includegraphics[width=16cm]{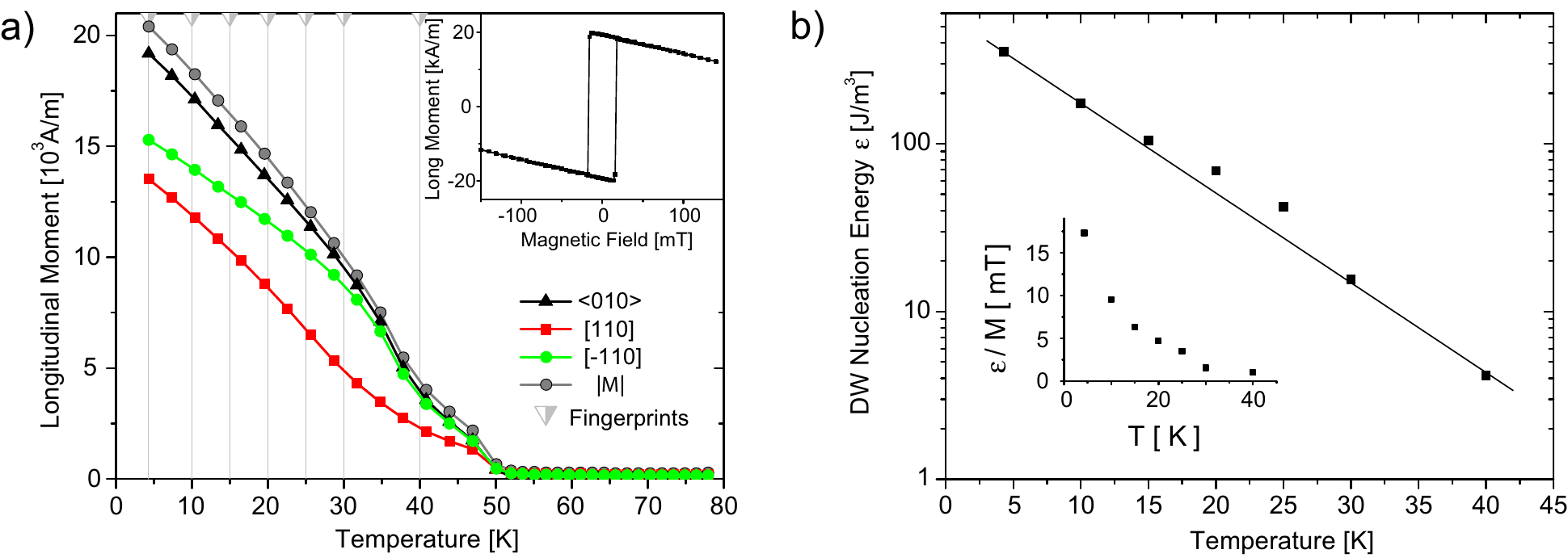}
\caption{\label{fig:FP-SQUIDeps} \textit{ a) Measurement of the
projection of the remanent magnetic moment of the sample of
Figs.~\ref{fig:FP-1} and \ref{fig:FP-2} onto different crystal
axes by SQUID magnetometry. Vertical gray lines indicate the
temperatures of the fingerprint measurements in
Figs.~\ref{fig:FP-1} and~\ref{fig:FP-2}. b) Domain wall nucleation
energy $\varepsilon_{90^\circ}$ (symbols) versus temperature,
derived from $\varepsilon_{90^\circ}/M$ (inset) extracted from the
fingerprints. } }
\end{figure}

Fig.~\ref{fig:FP-SQUIDeps}a shows SQUID measurements on the sample
of Figs.~\ref{fig:FP-1} and \ref{fig:FP-2}. After magnetizing the
sample along a given direction, we measure the projection of the
remanent magnetization on the respective axis and its evolution
with increasing temperature. Displayed are measurements along the
two 4 K hard magnetic axes [110] and $[\overline{1}10]$ and one of
the biaxial easy axes $\langle 010 \rangle$. They show the same
anisotropy transition as the fingerprints above. At 4 K, the
$\langle 010 \rangle$ crystal direction is close to a global
magnetic easy axis and thus shows the largest projection of the
remanent magnetic moment. The $[\overline{1}10]$ direction
coincides with the easy axis of the uniaxial
K$_{uni[\overline{1}10]}$ anisotropy term. That is why it is
closer to a global easy axis than the [110] direction \cite{note}
and in consequence shows a larger projection of the remanent
moment. As temperature increases, the magnetization decreases and
the relative amplitude of the anisotropy terms changes, as
described above. This results in a gradual reorientation of the
global easy axes with temperature, changing the angle between
remanent magnetization (along the global easy axis closest to the
sweep direction in Fig.~\ref{fig:FP-SQUIDeps}a) and the respective
sweep direction. The result of both the decreasing volume
magnetization and changing relative projections onto the different
sweep directions, can be seen in Fig.~\ref{fig:FP-SQUIDeps}a. The
green $[\overline{1}10]$ remanence, e.g., gains relative weight
with increasing temperature. This supports the observations of the
fingerprint measurements, where the $[\overline{1}10]$ anisotropy
term gains in influence at higher temperatures. Given the specific
anisotropy behaviour, known from the transport measurements, we
can estimate the absolute value of the remanent magnetization from
the square root of the sum of the squares of the two magnetization
projections along $[\overline{1}10]$ and [110](Pythagorean
theorem) \cite{Mike05PRL}. The result is displayed in gray in
Fig.~\ref{fig:FP-SQUIDeps}a. Such a magnetization measurement with
SQUID is complementary to transport investigations, since those
can only give energy scales in field units, i.e. normalized to the
volume magnetization like K/$M$ or $\varepsilon$/$M$.

The quantitative determination of the anisotropy constants at
higher temperatures is more complex than at 4 K and work is
ongoing to find a set of straightforward rules as for the mainly
biaxial system at 4 K. Determining the domain wall
nucleation/propagation energy $\varepsilon$, however, is possible
with the described techniques. Black symbols in
Fig.~\ref{fig:FP-SQUIDeps}b show preliminary results determined
from the fingerprints in Figs.~\ref{fig:FP-1} and ~\ref{fig:FP-2}.
The line is a guide to the eye. The method for the extraction
builds on the techniques described in
section~\ref{sec:el-biaxial}: 2$\varepsilon$/$M$ is basically
given by the diagonal of the rectangular first switching field
pattern for mainly biaxial samples and by the easy axis direction
diameter for purely uniaxial samples. The strength of this method
is that we can extract $\varepsilon_{90^\circ}$ easily from the
plots, because the global easy axes directions are obvious from
symmetry considerations. It is not necessary to assume a constant
(or known) global easy axis direction and we can thus fully
account for the complex temperature dependence of the easy axis
behaviour without fitting the data to a complicated model. Both
the determination of M and of $\varepsilon$/$M$ are not as
accurate in the transition region, where the energy surface at
zero field is almost flat over a wide angular range. This is a
probable cause of the deviation from perfect exponential behaviour
for the data in Fig.~\ref{fig:FP-SQUIDeps}b at intermediate
temperatures.

The square hysteresis loop with abrupt switching events, shown in
the inset of Fig.~\ref{fig:FP-SQUIDeps}a, points to a DW nucleation
dominated process, as opposed to a process, where the energy needed
for DW propagation is the limiting parameter \cite{Ferre}. Also the
temperature dependence of the DW nucleation energy in
Fig.~\ref{fig:FP-SQUIDeps}b fits to the standard exponential
behaviour expected for the temperature dependence of the coercivity
\cite{note2,exponCoersivity}. We suggest that the above method is
one tool that, in combination with, e.g., time dependent and optical
investigations \cite{Thevenard}, can clarify the DW nucleation
process in (Ga,Mn)As. It can complement recent optical studies, that
identify the nature of pinning centers and visualize the process of
DW-related magnetization switching in (Ga,Mn)As \cite{WangPinning,
Dourlat}.

\begin{figure}[tb]
\centering
\includegraphics[width=16cm]{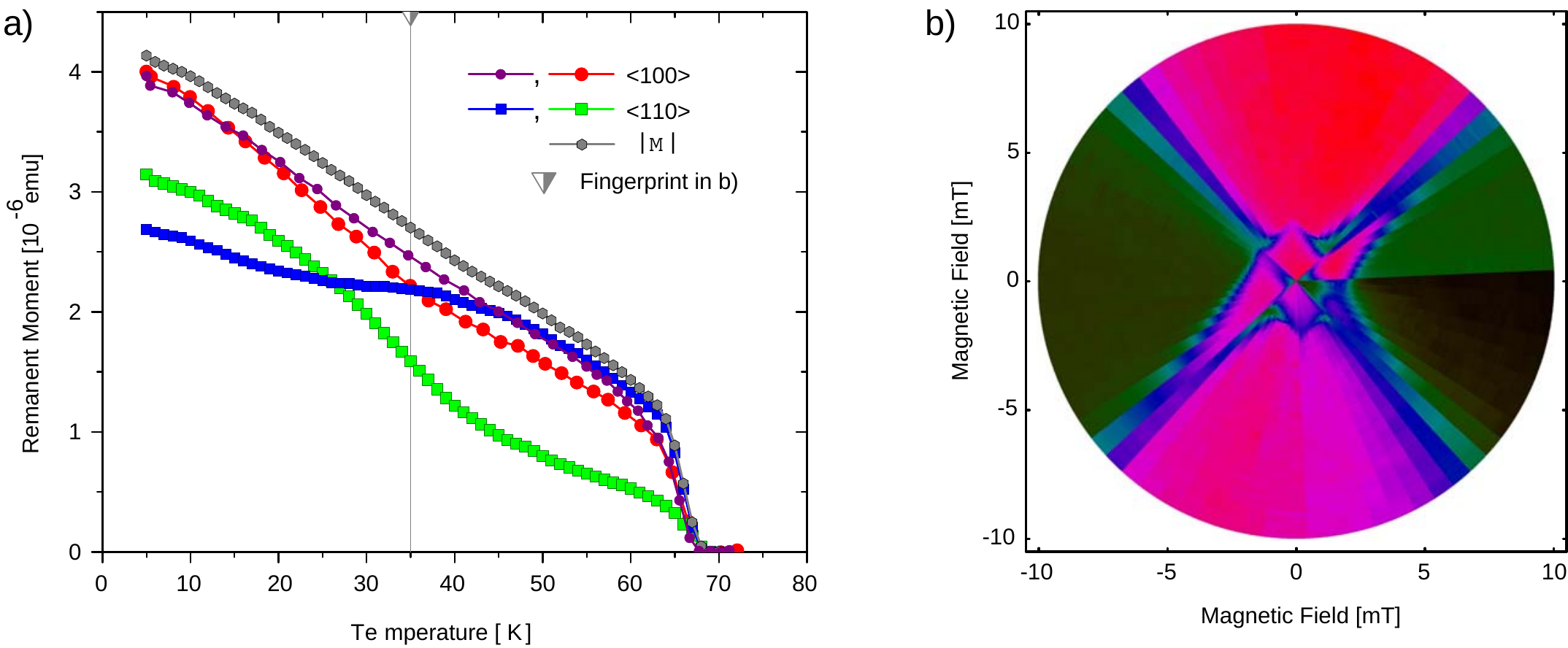}
\caption{\label{fig:FP-35K} \textit{a) Projection of the remanent
magnetic moment of a 20 nm thick (Ga,Mn)As layer measured with
SQUID along different crystal axes. 35 K, the temperature of the
anisotropy fingerprint measurement \cite{note3} in b) is indicated
by a vertical gray line.} }
\end{figure}

SQUID studies on another sample are shown in
Fig.~\ref{fig:FP-35K}a. As in Fig.~\ref{fig:FP-SQUIDeps}a we plot
the projection of the remanent magnetic moment vs temperature.
Shown are measurements along the two biaxial easy axes (red and
purple) and along the two bisecting directions (green and blue).
The absolute value of the volume magnetization is estimated as
discussed above (gray). The large difference between the two
biaxial easy axes directions (red and purple) at intermediate
temperatures (25 to 60 K) points to a symmetry breaking caused by
a relatively strong uniaxial [010] anisotropy component. For this
reason we investigate this sample at 35 K, where the [010]
component should be strongly visible in the symmetry of the
fingerprint pattern, and where the transport signal is still large
enough to get clean measurements. Fig.~\ref{fig:FP-35K}b shows the
resulting fingerprint. The symmetry breaking between the two
biaxial easy axes (here along 0$^\circ$ and 90$^\circ$) is
apparent from the picture. The relatively strong uniaxial [010]
term causes a preference for the magnetization orientation along
$90^\circ$. The resistance polar plot in turn resembles in parts a
typical biaxial fingerprint pattern (between 45$^\circ$ and
135$^\circ$ and the point symmetric region) and in the other
quadrants a typical uniaxial fingerprint pattern (between
135$^\circ$ and 225$^\circ$).\cite{note3} We can thus conclude,
that the relatively small uniaxial term gains in importance at
intermediate temperatures in this sample. This is where the two
stronger anisotropy terms have approximately equal weight,
compensating each other in specific angular regions. A small extra
term in the energy equation then plays a huge role: it creates an
additional local minimum in the energy surface, causing very
different switching behaviour in different quadrants of the polar
plot.

In summary, we have shown that the extended anisotropy fingerprint
technique is a powerful method to access the fine details of
complex anisotropies in ferromagnetic semiconductors. We used this
method to show that all transport layers investigated showed three
symmetry components of the magnetic anisotropy; the main biaxial
term and two uniaxial terms along the $[\overline{1}10]$ and the
[010] crystal directions. The relative strength of these
anisotropy terms is roughly speaking, of the order of
$K_{cyst}:K_{uni[\overline{1}10]}:K_{uni[010]}\sim 100:10:1$ at
4~K. At higher temperatures the relative strength of the
$[\overline{1}10]$ anisotropy component increases. The overall
behaviour of the anisotropy terms is consistent with SQUID
investigations, showing the typical transition from a mainly
biaxial to a mainly uniaxial material with increasing temperature.
An extraction of the 90$^\circ$-DW nucleation energy and its
temperature dependence is also possible. Measurements have shown,
that the [010] uniaxial anisotropy term, whose existence is
sometimes questioned, can be clearly observed. We show that it can
have a particularly strong impact on the switching behaviour for
cases where the cooperative effect of the biaxial and the
$[\overline{1}10]$ uniaxial anisotropy term lead to a flattened
energy surface.

\section*{Acknowledgements}
The authors wish to thank S. H\"{u}mpfner and V. Hock for sample
preparation and C. Chappert and W. Van Roy for useful discussions,
and acknowledge the financial support from the EU (NANOSPIN
FP6-IST-015728)and the German DFG (BR1960/2-2).


\begin{thebibliography}{99}
\bibitem{note} See Fig.~\ref{fig:110lines}, where [110] is along
$45^\circ$ and $[\overline{1}10]$ along $135^\circ$.
\bibitem{note2} The extraction method already accounts for the correction of
the temperature dependence of the anisotropy and the understanding
of the respective fingerprint pattern guaranties that only
switching events due to DW nucleation are analyzed, as distinct
from switching events that are caused by Stoner-Wohlfarth
rotation. The "constant anisotropy" criterion for the exponential
behaviour is thus satisfied.
\bibitem{note3}
The small resistance deviation of $\sim0.7$\% in the last quadrant
is caused by temperature drift. The temperature decreased by
$\sim1 K$ during the measurement between $\sim300^\circ$ and
$360^\circ$, changing the relative weight of the biaxial and the
uniaxial anisotropy components, slightly modifying the fingerprint
pattern in this section.
\end{thebibliography}
\end{document}